\documentclass[letterpaper,twocolumn,10pt]{article}
\usepackage{usenix}
\usepackage{tikz}
\usepackage{amsmath}
\usepackage{tabularx}

\usepackage{enumitem}
\usepackage{url}

\usepackage{pifont}

\usepackage{listings}
\lstdefinelanguage{CapnProto}
{
  morekeywords={struct,union,enum},
  keywordstyle=\color{blue},
  identifierstyle=\color{teal},
  stringstyle=\color{red},
  basicstyle=\tiny\ttfamily,
  commentstyle=\color{green}\ttfamily,
}

\usepackage{tcolorbox}
\tcbuselibrary{listings,skins,breakable}
\tcbuselibrary{listings}
\definecolor{OliveGreen}{rgb}{0.3,0.7,0.4}
\definecolor{crimson}{RGB}{220,20,60}
\definecolor{ngreen}{RGB}{150,195,125}
\definecolor{nyellow}{RGB}{243,210,102}
\definecolor{nred}{RGB}{216,56,58}
\definecolor{darkred}{rgb}{0.5, 0.0, 0.0}
\definecolor{grey95}{gray}{0.95}

\newtcolorbox{takeaway}{colframe=black,colback=gray!15,boxrule=1pt,arc=2pt,left=2pt,right=2pt,top=1pt,bottom=1pt, before skip=1em, after skip=0em}
\newtcbox{\inlinecolorbox}[1][white]{%
  on line,
  arc=0pt,
  outer arc=0pt,
  colback=#1,
  colframe=#1,
  boxrule=0pt,
  boxsep=0pt,
  left=0pt,
  right=0pt,
  top=0pt,
  bottom=0pt,
  fontupper=\ttfamily,
  nobeforeafter,
  hyphenationfix,
  breakable,
}
\usepackage{fvextra}
\DefineVerbatimEnvironment{verbatim}{Verbatim}{breaklines=true}

\usepackage{enumitem}
\usepackage{pifont}    
\definecolor{DarkGreen}{RGB}{0,215,0}
\definecolor{amethyst}{rgb}{0.6, 0.4, 0.8}
\definecolor{DarkYellow}{RGB}{230,230,0}
\definecolor{DarkRed}{RGB}{215,0,0}

\newcommand{\greencheck}{{\color{DarkGreen}\ding{52}}}   
\newcommand{\redcross}{{\color{DarkRed}\ding{56}}}       

\usepackage{amssymb} 
\usepackage{fontawesome5}

\newcommand{\exploitable}{\textcolor{DarkRed}{\faBolt}\textcolor{DarkRed}{\textsuperscript{*}}}
\newcommand{\highrisk}{\textcolor{DarkRed}{\faBolt}}
\newcommand{\notexp}{\texttt{[x]}}
\newcommand{\na}{\emptcirc{}} 

\usepackage[graphicx]{realboxes}
\usepackage{float}
\usepackage{tikz}
\newcommand*\emptcirc[1][1ex]{\tikz\draw (0,0) circle (#1);} 

\newcommand*\fullcirc[1][1ex]{\tikz\fill (0,0) circle (#1);} 

\usepackage{hyperref}

\newtcolorbox{mybox}{ colframe=black,colback=gray!15,boxrule=1pt,arc=2pt,left=2pt,right=2pt,top=1pt,bottom=1pt}

\usepackage[table]{xcolor}
\usepackage{diagbox}
\usepackage{booktabs}
\usepackage{multirow}
\usepackage{graphicx}
\usepackage{subcaption}
\usepackage{makecell}
\newif\ifcommentcond
\commentcondtrue 

\newif\ifupdatecond
\updatecondfalse 

\newcounter{wqy} 
\newcounter{todo}

\newif\ifparasumcond
\parasumcondfalse 

\newcommand{\ignore}[1]{}

\newcommand{\sysname}{\textsc{x402scope}}

\newcommand{\ruleNum}{eight}
\newcommand{\violateruleNum}{49}
\newcommand{\bugClass}{four}
\newcommand{\bugCount}{31}  

\newcommand{\facilitatorNum}{15}

\begin{document}

\date{}

\title{When HTTP 402 Meets the Blockchain: Risks on Emerging x402 Payments}

\newcommand{\equalcontrib}{\textsuperscript{\normalfont *}}
\newcommand{\correspondingauthor}{\textsuperscript{\normalfont\ding{41}}}

\newcolumntype{C}{>{\centering\arraybackslash}X}
\author{
\begin{tabularx}{0.92\textwidth}{CCC}
{\rm Qinying Wang}\equalcontrib & {\rm Yong Yang}\equalcontrib & {\rm Yuan Chen}\\
EPFL & Zhejiang University & Independent Researcher
\end{tabularx}\\[2ex]
\begin{tabularx}{0.62\textwidth}{CC}
{\rm Shouling Ji}\correspondingauthor & {\rm Mathias Payer}\\
Zhejiang University & EPFL
\end{tabularx}
} 

\maketitle
\begingroup
\renewcommand{\thefootnote}{\fnsymbol{footnote}}
\footnotetext[1]{The authors contributed equally to this work.}
\endgroup
\begingroup
\renewcommand{\thefootnote}{\protect\ding{41}}
\footnotetext{Corresponding author.}
\endgroup

\begin{abstract}
x402 is an emerging payment protocol for Web APIs and autonomous AI agents.
It is driven by the rise of LLM-based agents that can autonomously purchase access to online services.
x402 extends HTTP 402 with a payment negotiation flow and delegates payment proof verification and on-chain settlement to third-party facilitators.
As a result, facilitators serve as a shared payment infrastructure for many independent merchants. This centralizes trust and validation in one component, so a single flaw can affect many services.
Despite rapid adoption by major vendors and economically meaningful mainnet activity, the security posture of real-world x402 deployments remains poorly characterized.

We present the first systematic study of authorization correctness and execution safety in current facilitator-mediated x402 deployments in the wild, identifying \ruleNum{} security rules for facilitators as critical payment infrastructure.
Based on our analysis of rule violations, we derive four new attack
vectors, including \emph{Free Shopping}, \emph{Asset Theft}, \emph{Service Denial}, and \emph{Gas Abuse}.
These attacks exploit weaknesses in the real-world facilitator and server 
implementations and cause severe harm, including direct financial 
loss to merchants, theft of facilitator-held assets, unbounded 
sponsor-paid gas/fees, and disruption of payment services.
To assess the security of x402 deployments at scale, we propose a 
semi-automated black-box tool and apply it to \facilitatorNum{} major
x402 facilitators collectively used by over 60K sellers and 360K buyers.
Alarmingly, we find violations in all evaluated facilitators.
We responsibly disclosed 
our findings to the affected parties, who acknowledged the issues and adopted mitigations, including changes by Coinbase.
Finally, we complement our controlled testing with an empirical measurement of over 119 million recent Base and Solana transactions, quantifying x402 adoption, facilitator centralization, and ecosystem-level risk indicators.
\end{abstract}


\section{Introduction}
x402 is an emerging payment protocol for Web APIs and autonomous AI agents, motivated by the growing deployment of LLM-based agents that can discover, pay for, and consume web services programmatically~\cite{birch2025agentic, rothschild2025agentic, sapkota2025ai}.
Unlike traditional card payments that are designed around interactive user sessions and delayed reconciliation, x402 offers an API-native, programmable paywall for near-instant micropayments with enforceable policy constraints.
This design has enabled rapid ecosystem uptake.
x402 is already deployed by major vendors, including Coinbase~\cite{coinbase_x402}, Cloudflare~\cite{cloudflare_agents}, AWS~\cite{aws_builder_x402}, Circle~\cite{circle_blog_x402}, and Google~\cite{google_a2a_x402}. 
x402 has also reached economically meaningful mainnet activity. x402scan reports over 150M on-chain transactions and over \$40M in cumulative volume, involving over 400K buyers and over 80K sellers~\cite{x402scan}.

Technically, x402 extends HTTP 402 (``Payment Required'') with a payment negotiation flow and uses blockchains such as Base and Solana~\cite{yakovenko2018solana} for settlement.
The specification defines three roles: \emph{clients} (customers), \emph{servers} (merchants), and \emph{facilitators} (trust-bearing intermediaries). 
A client attaches proof of payment to a request. 
The server specifies the payment requirements and forwards the proof to a facilitator for verification and on-chain settlement.
The facilitator returns verification and settlement responses to the server, which uses them to gate access to protected endpoints.


Despite this rapid adoption, the security posture of real-world x402 deployments remains poorly characterized, particularly for facilitators. As shared payment infrastructure provides services for many independent applications, facilitators concentrate trust and validation logic in a single component, increasing the blast radius of failures. A flawed or compromised facilitator may misauthorize access, misdirect payments, or turn settlement into an attacker-controlled cost sink across multiple services.
This can cause direct financial loss and cascading disruption across the ecosystem, impacting a large number of servers and clients.
Prior work has primarily focused on client-side threats, including Sybil-based service 
discovery for x402~\cite{shi2025sybil}, secure mechanisms for delegating payment 
permissions to agents~\cite{teamhuman}, and safeguards to reduce the risk of agents 
being induced or mistakenly initiating or settling payments~\cite{x402secure}.
However, the security of facilitators' payment verification and on-chain settlement in real deployments
remains underexplored.
Thus, a pressing question arises:
\emph{How should we reason about facilitators' security in x402
deployments and what is the resulting security impact?}
To address this, we identify two key challenges:

\noindent\textbf{Challenge 1. Facilitator semantics gaps between layers and networks.}
Reasoning about facilitator security is hard because facilitators sit at the boundary of multiple systems 
with mismatched semantics and timing assumptions.
In practice, the facilitator translates the web-layer payment proofs into concrete 
verification decisions and on-chain settlements.
It must bridge HTTP request semantics, validation logic, and blockchain execution, 
which differ in trust and timing assumptions.
Moreover, x402 deployments span networks with substantially different payment proof formats and execution semantics, including chain-specific account and rent mechanics.
Achieving authorization correctness requires understanding how fields, checks, and failure modes compose across layers and networks.
This makes correctness reasoning difficult to capture with single-layer analyses or ad hoc validations.

\noindent\textbf{Challenge 2. Facilitator customization and limited observability.}
Even if we know which security properties should hold, assessing them in the wild is difficult because real-world facilitators are \emph{heavily customized and largely black box}.
Many are closed source and only exposed via remote APIs, so we cannot rely on code-centric analysis to recover their effective checks, supported payment proof types, or failure handling logic.
Moreover, facilitators often differ in their enabled features, such as supported networks, proof types, verification, and settlement semantics. 
This feature heterogeneity complicates systematic evaluation because the same observed behavior may arise from different feature configurations.

\noindent\textbf{Solution.}
To handle Challenge 1, we translate the x402 workflow into a small set of checkable security rules.
Specifically, we systematically analyze the x402 workflow and enumerate the supported payment proof types and their semantics.
Based on this end-to-end analysis, we distill a set of security rules that capture the necessary invariants covering authorization correctness and execution safety.
These rules provide a unified basis for understanding security risks
across diverse x402 deployments, rather than ad hoc analysis of individual implementations.
To address Challenge 2, we develop \sysname{}, a semi-automated black-box testing tool that assesses the security of heterogeneous facilitators at scale.
Leveraging the unified rules, \sysname{} adopts a feature-aware, rule-guided approach~\cite{chen2025unveiling}.
It first performs capability discovery to infer each facilitator’s enabled features, such as supported networks and payment proof types, and then generates only applicable tests.
Concretely, \sysname{} maintains payment-proof templates across networks and proof formats, then it mutates them according to the security rules and discovered capabilities.
Using HTTP responses and on-chain transaction receipts as oracles, \sysname{} identifies violated rules and their corresponding attack implications.
Importantly, \sysname{} requires no source code, facilitator private keys, or internal state, enabling evaluation in the wild.

Based on rule violations, we reveal \bugClass{} new attack classes.
\begin{itemize}[itemsep=0pt, parsep=0pt, topsep=0pt]
  \item \textbf{Free shopping attack} allows an attacker to obtain 
  resources without paying by exploiting payment-ordering flaws, 
  timing windows, or flawed payment verification.
\item \textbf{Asset theft attack} extracts funds by exploiting flaws in
payment proof verification or settlement, diverting assets 
to attacker-controlled accounts.
\item \textbf{Service denial attack} renders the server unable to
provide service, while the client completes payment without receiving
the requested resources.
\item \textbf{Gas abuse attack} allows an attacker to 
force the facilitator to pay gas for attacker-controlled
deployments or unbounded execution.
\end{itemize}

\noindent\textbf{Evaluation.}
Using \sysname{}, we present the first security study of \facilitatorNum{} real-world facilitators that are used by over 60K sellers and
360K buyers and collectively account for 99\% of x402 transactions.
We identify \violateruleNum{} security rule violations, which translate into \bugCount{} previously unknown vulnerabilities. 
We find systematic non-compliance in practice: every evaluated facilitator violates at least one rule, and every rule is violated by at least one platform. The dominant practical risks are sponsor-paid cost amplification and free shopping, while asset theft is less frequent but has the highest impact, including cases in top volume facilitators. We responsibly disclosed all issues to the respective maintainers, who acknowledged our findings and provided mitigations, including Coinbase.
We further complement our controlled testing with a large-scale measurement of 119 million x402-related transactions on Base and Solana, characterizing adoption, centralization, and ecosystem-level risk indicators.
Because historical attacks lack ground truth, we treat this measurement as risk evidence rather than attack attribution. 
The results show sustained usage and non-trivial settlement failure rates.
In total, x402-related settlement attempts have already burned over \$202K in gas and fees, including \$5.8K from reverting submissions alone, demonstrating direct sponsor-side loss. 
We observe over 22.9K transactions with ATA-creation patterns that cost more than \$5.7k, highlighting ecosystem-level exposure to costs paid from facilitator-held assets.

\noindent\textbf{Our contributions.}

\begin{itemize}[itemsep=0pt, parsep=0pt, topsep=0pt]
  \item We present the first security analysis of real-world x402 facilitators and distill \ruleNum{} unified security rules. Guided by these rules, we identify \bugClass{} practical attack classes.
  \item We build a semi-automated black box testing tool and apply it to \facilitatorNum{} major facilitators, uncovering \bugCount{} previously unknown vulnerabilities that we responsibly disclosed to maintainers. Our results show that security rule non-compliance is widespread in deployed x402 systems and yields high-impact failures in verification and settlement.
  \item We conduct a large-scale on-chain measurement study to characterize x402 adoption, facilitator centralization, and ecosystem-level risk indicators.
\end{itemize}










\section{Background}
\label{sec:back}
\subsection{Overview of x402 Protocols}
\label{sec:overviewofx402}

\begin{figure}[ht]
  \centering
  \includegraphics[width=\linewidth]{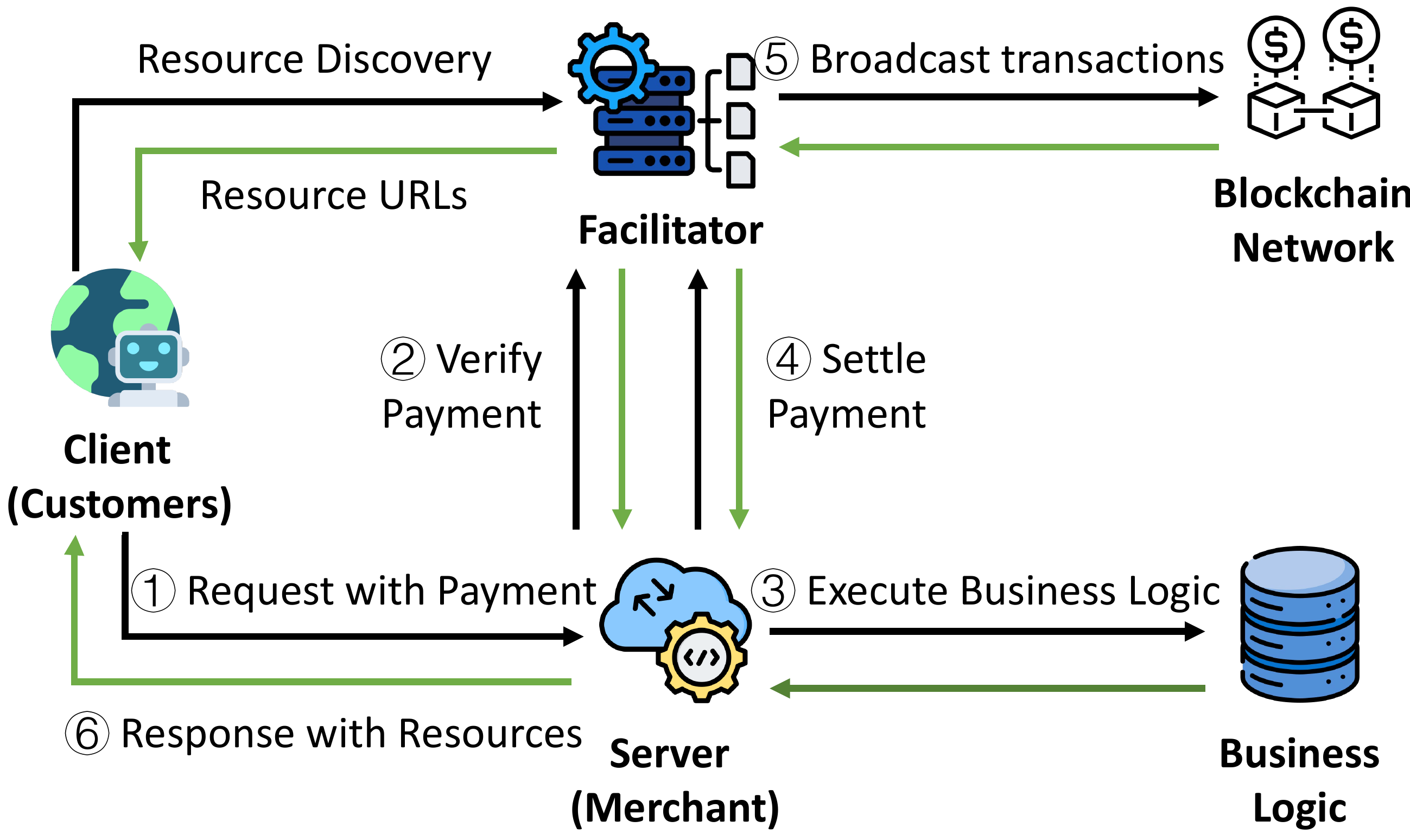} 
  \caption{Overview of x402 protocol payment paradigms.}
  \label{fig:x402diagram} 
\end{figure}

Building on the x402 protocol, an ecosystem is emerging around wallets, checkout clients, and server-side deployments (e.g., content providers) for paywalled endpoints and pay-per-request applications.
To illustrate this emerging architecture, \autoref{fig:x402diagram} shows an overview of a standard x402 payment.
Before initiating the payment-and-access loop, the client may perform resource discovery via the facilitator to obtain the 
URLs (or entry points) of paid resources, and then use those URLs to proceed 
with the subsequent x402 payment flow.

\textbf{\ding{172} Request with payment.}
When a client requests a protected resource, the server returns HTTP 402 (Payment Required) with a payment challenge specifying the required terms (e.g., network, asset, amount, recipient).
The client then resubmits the request with a payment payload in the \emph{X-Payment} header shown below, which carries the signature and metadata needed for verification and settlement.
EVM- and Solana-specific payload structures are shown in~\autoref{sec:payment_payload}.
\begin{tcblisting}{
  listing engine=listings,    % change engine to listings
  listing only,
  listing options={          % 
    language=bash,
    basicstyle=\footnotesize\ttfamily,
    breaklines=true,
    columns=flexible,
    postbreak=\mbox{\textcolor{red}{$\hookrightarrow$}\space},
    emph={GET, HTTP/1.1},
    emphstyle=\color{blue}\bfseries
  },
  boxrule=0.1mm,top=0mm,bottom=0mm,
  colback=blue!5!white,colframe=blue!75!black,
  left=0mm
}
GET /paywalled-endpoint HTTP/1.1
Host: <Resource URL>
X-Payment: {
  "x402Version": 1,
  "scheme": "exact",
  "network": "base",
  "payload": { <PAYMENT_PAYLOAD> }}
\end{tcblisting}



\textbf{\ding{173} Verify payment.} Upon receiving a request with an \emph{X-Payment} header, the server decodes it into the x402 payment.
The server then sends the payload and the payment requirements to a verification endpoint (e.g., \texttt{/verify}) as shown below. 
The facilitator validates the proof off-chain against the declared requirements, including a sufficient amount, a consistent recipient, and a well-formed proof. The facilitator returns a verification response, and the server uses it to gate access to the client's request.
Payment proofs vary in format and semantics across networks, so facilitators must perform chain-specific validation. Otherwise, they may accept malformed or misinterpreted proofs that bypass payment.
\begin{tcblisting}{
  listing engine=listings,    % change engine to listings
  listing only,
  listing options={          % 
    language=bash,
    basicstyle=\footnotesize\ttfamily,
    breaklines=true,
    columns=flexible,
    postbreak=\mbox{\textcolor{red}{$\hookrightarrow$}\space},
    emph={GET, HTTP/1.1},
    emphstyle=\color{blue}\bfseries
  },
  boxrule=0.1mm,top=0mm,bottom=0mm,
  colback=blue!5!white,colframe=blue!75!black,
  left=0mm
}
POST https://<facilitator url>/verify
{ "paymentPayload": <PAYMENT_PAYLOAD_FROM_CLIENT>,
  "paymentRequirements": {
    "scheme": "exact",
    "network": "base",
    "asset": <TOKEN_CONTRACT_ADDRESS>
    "payTo": <SERVER_PUBKEY>, // Server Wallet Address
    "maxAmountRequired": 1000
    "resource": <RESOURCE_URL>
    "maxTimeoutSeconds": 60 }}
\end{tcblisting}
\textbf{\ding{174} Execute business logic.}
After the payment proof is verified, the server triggers the downstream business workflow (e.g., preparing the resource, placing an order, or updating database).
Once the workflow completes, the business component reports the execution result to the server.

\textbf{\ding{175}\ding{176}  Settle payment and broadcast a transaction.} 
After executing business logic, the server delegates on-chain settlement to the facilitator.
It sends the same payment payload with the payment requirements to a settlement endpoint (e.g., \texttt{/settle}), and waits for the settlement outcome.
The facilitator then constructs and broadcasts the corresponding settlement transaction, pays the required gas/fees, and returns the transaction hash and status to the server.

\textbf{\ding{177} Respond with resources.} 
Once payment is confirmed, the server returns 
service response to the client.


\begin{tcblisting}{
  listing engine=listings,    % change engine to listings
  listing only,
  listing options={          % 
    language=bash,
    basicstyle=\footnotesize\ttfamily,
    breaklines=true,
    columns=flexible,
    postbreak=\mbox{\textcolor{red}{$\hookrightarrow$}\space},
    emph={GET, HTTP/1.1},
    emphstyle=\color{blue}\bfseries
  },
  boxrule=0.1mm,top=0mm,bottom=0mm,
  colback=blue!5!white,colframe=blue!75!black,
  left=0mm
}
HTTP/1.1 200 OK
X-PAYMENT-RESPONSE: {
  "success": true,
  "errorReason": null,
  "transaction": "0x2b28\ldots461e", // Transaction Hash
  "network": "base",
  "payer": <CLIENT_ADDRESS>} // Client Wallet Address
<RESOURCE_GENERATED_BY_BUSINESS_LOGIC>
\end{tcblisting}

\subsection{Payment Payload}
\label{sec:payment_payload}
The payment payloads are chain-dependent.
On Base and other Ethereum Virtual Machine (EVM)-compatible networks, the payload is an ERC-3009 \texttt{transferWithAuthorization}
proof \cite{erc3009}, as shown below.
The \texttt{signature} field can be an EIP-712 typed data
signature \cite{eip712} or an ERC-1271 smart-wallet signature \cite{erc1271}, optionally wrapped with ERC-6492 for undeployed wallets \cite{erc6492}.
It lets a client authorize a token transfer via an off-chain signature that the facilitator can later submit on-chain.
These variants place different validation logic on the facilitator-side 
workflow, which we analyze in~\autoref{sec:securityanalysis}.

\begin{tcblisting}{
  listing engine=listings,    % change engine to listings
  listing only,
  listing options={          % 
    language=bash,
    basicstyle=\footnotesize\ttfamily,
    breaklines=true,
    columns=flexible,
    postbreak=\mbox{\textcolor{red}{$\hookrightarrow$}\space},
    emph={GET, HTTP/1.1},
    emphstyle=\color{blue}\bfseries
  },
  boxrule=0.1mm,top=0mm,bottom=0mm,
  colback=blue!5!white,colframe=blue!75!black,
  left=0mm
}
{ "x402Version": 1,
  "scheme": "exact",
  "network": "base",
  "payload": { "signature": <Authorization Signature>,
    "authorization": {
      "from": <CLIENT_PUBKEY>,  // Client Wallet Address
      "to": <SERVER_PUBKEY>,    // Server Wallet Address
      "value": 1000,  // Payment Amount, 0.001 USDC
      "validAfter": 1768841386,  // Unix Timestamp
      "validBefore": 1768841492, // Unix Timestamp
      "nonce": "0x33b4\ldots d9e4" } } }
\end{tcblisting}

On the Solana Virtual Machine (SVM), x402 relies on native transaction semantics: the payload is
a serialized SPL Token or Token-2022 transfer
transaction (e.g., \texttt{transfer\_checked}), authenticated via Solana's Ed25519 transaction signature model.
As shown below, it carries a base64-encoded byte string containing the wire-format bytes of a Solana \texttt{VersionedTransaction} (message v0). We illustrate the decoded transaction semantics below. 
\begin{tcblisting}{
  listing engine=listings,    % change engine to listings
  listing only,
  listing options={          % 
    language=bash,
    basicstyle=\footnotesize\ttfamily,
    breaklines=true,
    columns=flexible,
    postbreak=\mbox{\textcolor{red}{$\hookrightarrow$}\space},
    emph={GET, HTTP/1.1},
    emphstyle=\color{blue}\bfseries
  },
  boxrule=0.1mm,top=0mm,bottom=0mm,
  colback=blue!5!white,colframe=blue!75!black,
  left=0mm
}
{ "x402Version":1,
 "scheme": "exact",
 "network": "solana",
 "payload":{ "transaction": { 
  "type": "VersionedTransaction",
  "signatures": [<sig1>, <sig2>, \ldots ],
  "message": {
    "version": 0,
    "header": {
      "numRequiredSignatures": 1,
      "numReadonlySignedAccounts": 0,
      "numReadonlyUnsignedAccounts": 6},
    "accountKeys": [
      <FEEPAYER_PUBKEY>, // Provided by Facilitator
      <ComputeBudget_ID>, <SPLTOKEN_ID>, // Program IDs
      <CLIENT_ATA>, <SERVER_ATA>, // Token Accounts
      <CLIENT_PUBKEY>, <MINT_PUBKEY>],
    "recentBlockhash": <RECENT_BLOCKHASH_BASE58>,
    "instructions": [
    { "program": ComputeBudget,
      "op": setComputeUnitLimit
      "value": <gaslimit> },
    { "program": ComputeBudget, 
       "op": setComputeUnitPrice, 
       "value": <gasprice> },
    { "program": SPL Token,
      "op": transfer_checked,
      "mint": <Token Address>,
      "source": <Client Token Account>,
      "dest": <Server Token Account>,
      "owner": <Client Pubkey>,
      "amount": 1000,  // Payment Amount, 0.001 USDC
      "decimals": 6 } ],
    "addressTableLookups": [] }} 
    //In practice, the transaction is serialized to its binary form and base64-encoded.
  } }
\end{tcblisting}

The decoded transaction specifies the token mint, amount, source and
destination token accounts, and signing authority. 
These token accounts are often Associated Token Accounts (ATAs), i.e., canonical token accounts derived from a wallet address and a token mint. Creating an ATA requires a rent-exempt lamport balance, which acts as a storage deposit and can be reclaimed when the account is closed.

Other networks may adopt analogous designs using their native transaction formats and signature semantics (e.g., a signed Starknet \texttt{invoke} transaction
authenticated under their native signature scheme, with fees
sponsored by a paymaster).

\subsection{Comparison with Card Payments}
The x402 workflow separates verification and settlement, analogous to credit card
\emph{authorization (hold)} and \emph{capture}. In card payments, the issuer
authorizes a purchase by checking card validity, available credit, and fraud signals.
Then it (logically) places a hold that gives the merchant a guarantee window; later,
the merchant captures the final amount to transfer funds (with interbank settlement
often completed asynchronously).

However, x402 operates under a different trust and cost model. Once an on-chain
payment is settled, it is effectively irreversible and lacks chargeback/dispute-style
rollback mechanisms. Therefore, the server first uses verification to validate the
proof and gate execution, and only triggers settlement after the business logic
succeeds. Moreover, x402 provides no credit-based guarantee (no issuer-backed credit
line), and many payments are micro-transactions where per-transaction fees dominate; this also creates sponsor-paid fee abuse risks, making early rejection via verification
critical before spending on-chain resources in settlement.
The concrete formats of payment payloads are chain-dependent.

\section{x402 Security Analysis}
\label{sec:securityanalysis}
This section analyzes authorization correctness and execution safety in the facilitator-mediated x402 workflow and its supported payment proof formats, distilling the security rules that motivate our analysis tools.
x402 differs from traditional payments in both trust and cost.
Because settlement is effectively irreversible and x402 provides no credit-based guarantees, verification is the service gate and must strictly enforce \emph{authorization correctness}, including proof validity and the binding of proof requirements.
Moreover, x402 targets micro-transactions where the facilitator sponsors on-chain gas/fees, making \emph{execution safety} essential.
Without it, attackers can manipulate settlement semantics to cause asset theft, for example, by substituting the intended transfer or injecting extra instructions, and to amplify sponsor-paid costs.

\noindent\textbf{Rule derivation and scope.}
We derive SR1--SR8 through a workflow-driven analysis of facilitator-mediated x402 payments. 
Specifically, we first examine the x402 specification~\cite{coinbase_x402_spec} to identify the parties, trust boundaries, and adversarial capabilities that define our threat model  (\autoref{sec:threatmodel}). 
We then analyze the supported payment proof types and the facilitator's two-stage \texttt{verify}/\texttt{settle} workflow. 
For each stage, we identify facilitator-side invariants that are observable in black-box testing and necessary for either \emph{authorization correctness} (\autoref{sec:correctness}) or \emph{execution safety} (\autoref{sec:safety}). 
These invariants form our Security Rules (SRs).
SR1-SR4 cover authorization correctness: payment proofs must match server-declared requirements, payer authorization and balance must be valid, freshness constraints must be enforced, and success must not be reported unless the corresponding payment condition is satisfied. 
SR5-SR8 cover execution safety: facilitators must reject non-settleable or economically meaningless payments, bound sponsor-paid costs, revalidate time- and state-dependent conditions before settlement, and restrict on-chain execution to well-defined payment semantics. 
Together, these rules cover the facilitator-side checks exercised by the current x402 workflow and supported proof formats.
These rules are not intended to cover all possible failures in payment systems. 
They exclude malicious or colluding facilitators, wallet compromise, merchant business logic beyond the payment boundary, full protocol redesigns, and credit- or collateral-based settlement alternatives.

\subsection{Threat Model}
\label{sec:threatmodel}
We consider the facilitator as a trust-bearing intermediary that may be buggy, misconfigured, or non-compliant, and study how adversarial clients/servers can exploit such weaknesses.
Accordingly, we consider two realistic adversarial settings.
First, a malicious client sends arbitrary payment proofs to an honest server to obtain service without a valid payment or to abuse the facilitator's side behavior. 
The impact of this setting depends in part on the server's deployment-specific resource-release boundary: while Figure~\ref{fig:x402diagram} presents a general multi-party payment workflow, real servers may release resources after facilitator verification or only after successful settlement. 
We therefore state the relevant release assumption when analyzing attacks such as \emph{Free Shopping}. 
Second, an attacker impersonating a server that crafts payment requirements together with proofs to trick the facilitator into incorrect settlement, causing asset loss or service unavailability. Both settings are realistic because facilitators and servers are globally accessible. 
Finally, we do not model a fully malicious or colluding facilitator as the primary adversary, as such a facilitator represents a trust-failure model rather than the protocol non-compliance problem targeted in this paper. We discuss this scope distinction in~\autoref{sec:discussion}.

\subsection{Authorization Correctness}
\label{sec:correctness}
Authorization correctness determines when a request is authorized as paid and eligible for access. In x402, correctness is non-trivial because proofs are chain-dependent and include network-specific fields. Facilitators must validate proofs precisely and ensure settlement matches the on-chain action implied by the server's payment requirements. We focus on Base and Solana, the two dominant x402 networks, which account for 97.13\% of observed transactions~\cite{hashed_official_x402_analytics}.

\noindent\textbf{What must be verified?}
First, the facilitator must strictly validate the server-declared
requirements and bind them to the payment payload semantics.
On Ethereum Virtual Machine (EVM), x402 payment payloads typically use ERC-3009 authorization transfers\cite{erc3009}, where the payer signs an off-chain authorization that specifies the recipient and amount and is later executed on chain. 
Accordingly, the facilitator must (i) enforce the declared requirements,
including \texttt{scheme}, \texttt{network}, \texttt{asset} (token contract),
\texttt{payTo}, and the required amount (e.g., \texttt{maxAmountRequired}), and
(ii) bind these requirements to the signed authorization by checking that
\texttt{payTo} matches the signed field \texttt{authorization.to}.
On Solana, x402 uses SPL Token transfers\cite{solana_transfer_tokens}.
In this setting, \texttt{asset} maps to the mint in the
\texttt{transfer\_checked} instruction, and the required amount maps to the
transfer \texttt{amount}.
Similarly, the effective recipient is the destination token account
\texttt{dest}, which must match the server's expected Associated Token Account
(ATA) derived from \texttt{payTo}.
Without these strict bindings, an adversary can exploit cross-version parsing
differences, cross-network confusion, or asset/recipient substitution, causing
the server to accept an invalid proof or a proof that is only valid in a
different context.
\begin{mybox}
\hypertarget{sr1}{\textbf{SR1.}}
During verification, a facilitator must return invalid if the payment 
  proof does not match the server-declared requirements. 
\end{mybox}

Second, the facilitator must validate payer authorization and
freshness.
On \textbf{EVM}, if the payer is an EOA, it verifies the EIP-712 signature \cite{eip712} over the
typed authorization message, enforces the validity window
(\texttt{validAfter}, \texttt{validBefore}), and ensures that the nonce has not
been used.
If the payer is a deployed contract wallet, it does not assume Elliptic Curve
Digital Signature Algorithm (ECDSA) semantics and instead validates the
signature by calling the wallet's on-chain signature checker (ERC-1271,
\texttt{isValidSignature}) \cite{erc1271}.
If the payer corresponds to a contract wallet that is not yet deployed, the
payment is accepted only with an ERC-6492-wrapped signature \cite{erc6492}.
The wrapper ensures that deployment to the intended address will succeed and
that the deployed contract will validate the same message/signature via
ERC-1271. This preserves an identical on-chain signature check at
settlement.
On \textbf{Solana}, the facilitator must verify that the transaction is signed
by the required client signer(s) and is still fresh.
A simulation can help detect missing or invalid signatures only if signature
verification is enabled.
Moreover, it checks \texttt{recentBlockhash} and rejects stale transactions. Otherwise, attackers can replay stale
authorizations or trigger a verification–settlement inconsistency, where a proof
passes verification but fails settlement.
\begin{mybox}
\hypertarget{sr2}{\textbf{SR2.}}
  During verification, a facilitator must return invalid if the payer 
  authorization is not authentic under the intended signature model.
\end{mybox}
\begin{mybox}
\hypertarget{sr3}{\textbf{SR3.}}
 During verification, a facilitator must return invalid if the payer 
  authorization is expired.
\end{mybox}
\noindent\textbf{What is considered paid?}
In x402 deployments, the server delegates payment settlement to a
facilitator and treats a valid response as sufficient evidence
to authorize access. 
However, ``paid'' is grounded in the on-chain settlement outcome: 
only a successfully settled transfer moves funds to the recipient.
If \texttt{valid} can be returned before such settlement succeeds during settlement, the
system introduces a semantic gap between authorization and payment.
An attacker can exploit this gap by submitting a proof that passes
verification but later reverts during on-chain settlement.
\begin{mybox}
\hypertarget{sr4}{\textbf{SR4.}}
 During settlement, a facilitator must return valid of a payment proof only when it is 
  settled on the chain.
\end{mybox}

\subsection{Execution Safety}
\label{sec:safety}
Motivated by x402's cost model, where facilitators often sponsor on-chain settlement costs such as gas and fees, we emphasize \emph{execution safety}. Execution safety specifies the on-chain actions a facilitator is allowed to perform during settlement and how sponsor funded execution can be abused.

\noindent\textbf{What attacker-controlled factors can increase the facilitator's
settlement cost?}
First, cost can increase when the facilitator submits transactions that are
destined to fail on chain.
On \textbf{EVM}, attacker-controlled proofs can trigger contract-level validation
failures (e.g., replayed nonces, nonce race, expired or about to expire validity windows, or insufficient
token balance), which revert after execution begins and still burn the sponsor paid
gas.
On \textbf{Solana}, crafted transactions can fail during program execution due to
invalid or mismatched account lists/authorities for the SPL Token
instruction, missing authorities, or insufficient funds, and execution
failures still charge the fee payer, even though the intended transfer does not
complete.

Second, the mechanism can be economically unviable even when the nominal payment amount is small or zero,
because each submission still incurs on-chain fees, and repeated zero-value requests can
cumulatively drain sponsor-paid fees.
Therefore, the facilitator must fail fast during verification by rejecting proofs
that are unlikely to settle, and by filtering out economically meaningless
payments, reducing unnecessary on-chain submissions and sponsor expenditure.
\begin{mybox}
\hypertarget{sr5}{\textbf{SR5.}}
During verification, a facilitator must fail fast by rejecting payment proofs
that are economically meaningless (e.g., zero-amount or an amount below a fee
threshold) or not settleable (e.g., insufficient balance), and must
enforce idempotency and a reasonable freshness bound.
\end{mybox}

Third, supporting ERC-1271 contract wallet signatures on EVM and SPL Token
transfers on Solana introduces attacker-controlled inputs that can amplify
sponsor-paid execution costs on each chain.
\textbf{On EVM}, ERC-1271 lets a contract account define its own signature
validation logic via \texttt{isValidSignature}.
ERC-6492 extends this model to
\emph{counterfactual} accounts by wrapping an ERC-1271-style contract signature
together with deployment or factory data.
If the facilitator sponsors settlement and processes attacker-supplied ERC-1271 or
ERC-6492 payloads, an attacker can amplify sponsor-paid gas by forcing expensive
\texttt{isValidSignature} execution and, under ERC-6492, additional factory calls
or contract deployment.
\textbf{On Solana}, an SPL token payment is realized by submitting a transaction that invokes the SPL Token program with an account list, instruction data,
and required signers.
Since the facilitator signs last and sponsors such transactions, attacker controlled
inputs can increase cost by inflating the account list and instruction footprint,
requiring extra signers, which increases
transaction size, compute usage, and signature verification overhead.
Without strict constraints and cost bounds, settlement can become sponsor-paid high-cost execution or repeated fee burning.
\begin{mybox}
\hypertarget{sr6}{\textbf{SR6.}}
Sponsored execution must be constrained by configurable upper bounds on 
fees, gas or compute units. 
\end{mybox}

Fourth, a proof may appear valid during verification but become invalid or unlikely to
settle by the time it is submitted.
This divergence is inherent to the x402 ``onion'' workflow: the facilitator first
verifies the client provided payment proof, then the server executes business
logic, and only after successful fulfillment does the facilitator attempt on-chain
settlement.
The resulting time gap introduces Time-of-Check to Time-of-Use (TOCTTOU) risks,
where freshness and feasibility assumptions may change between verification and
settlement.
For example, the authorization window may be near expiry, the recent blockhash
may become stale, or the payer balance may become insufficient.
Importantly, this is exacerbated by the fact that verification and settlement
are typically stateless and decoupled endpoints. Therefore, settlement should
re-validate all time- and state-sensitive constraints (or otherwise enforce a
binding verification context) rather than assuming that the earlier
verification result still holds.
\begin{mybox}
\hypertarget{sr7}{\textbf{SR7.}}
Before settlement submission, a facilitator should redo verification to avoid
unnecessary sponsor paid submissions that are destined to fail.
\end{mybox}

\noindent\textbf{Can a settlement proof be crafted to divert assets from the facilitator?}
When the facilitator signs and sponsors settlement transactions, attacker-controlled
payment payloads can steer on-chain execution toward unintended value flows.
\textbf{On EVM}, an attacker can embed arbitrary factory calldata in an ERC-6492
payload. If the facilitator sponsors settlement and executes it without enforcing
that it is a genuine deployment to the expected counterfactual address, the call
can instead run attacker-chosen logic that transfers ETH or tokens from the
facilitator to an attacker-controlled address.
\textbf{On Solana}, a crafted SPL token transaction can redirect value by
manipulating the instruction accounts and program invocations in the settlement
payload. In particular, the payload can force the facilitator, as the fee payer,
to fund and create an attacker-owned associated token account (ATA) during
settlement (e.g., via an injected ATA-creation step) and then route the transfer
to that account.
Therefore, the facilitator should bind settlement to an explicitly allow-listed
and unambiguous execution template (program IDs, instruction set, and expected
accounts) and reject any semantic deviation, so execution semantics remain
integral and predictable.

\begin{mybox}
\hypertarget{sr8}{\textbf{SR8.}}
A facilitator must settle only proofs whose on-chain execution semantics 
are explicitly allowed and unambiguous, rejecting any unexpected 
instructions, extra signers, or alternative token programs or contracts.
\end{mybox}
\section{New Attack Vectors}
\label{sec:attack}


Based on these assumptions and the preceding discussion on the security 
rules that facilitators must maintain, we identify four new attack vectors.

\noindent\textbf{Free shopping attack.}
Free shopping becomes possible when the server releases the protected resource
without a verifiably successful on-chain settlement.
We distill two root causes.
The first root cause is facilitator-side false settlement success.
If the facilitator violates \hyperlink{sr4}{\textbf{SR4}} and incorrectly
reports settlement success (e.g., returning \texttt{true} while the settlement
transaction reverts, is dropped, or is never confirmed on chain), then even a
server that gates delivery on the settlement response can be tricked into
releasing unpaid resources.
The second root cause is a server integration pitfall: the server executes
business logic after a successful verification response and may release the
protected resource before settlement is confirmed.
All seven official Coinbase reference server SDKs lack rollback for post-verification side effects. The official Flask SDK (v0.2.1) even releases the protected resource after verification, regardless of later settlement failure.

Under the second root cause (server-side verification-only acceptance), an
attacker can stably obtain unpaid service by inducing a \emph{verification-pass /
settlement-fail} divergence via the following mechanisms and rule violations:
(1) Concurrent verification (nonce race).
If the facilitator violates \hyperlink{sr5}{\textbf{SR5}} by failing to enforce
idempotency for repeated submissions of the same nonce-bound proof (i.e.,
missing deduplication or consistent outcome caching), an attacker can send
multiple parallel requests carrying the same payment proof.
At most one request
can be successfully settled and consume the nonce; the remaining requests become
non-settleable at settlement time and fail on chain. Under verification-only
acceptance, those failed settlements may still receive the protected service
because the server releases results after verification rather than after
confirmed settlement.
(2) Time-window TOCTTOU. 
If the facilitator violates \hyperlink{sr5}{\textbf{SR5}} by enforcing 
an overly permissive freshness bound or by not enforcing account-balance checks, 
an attacker can target short authorization windows or balance-dependent validity.
A proof can pass verification as fresh and sufficiently funded, 
but later expire or become underfunded by the time it is submitted for 
settlement or confirmed on chain.
(3) ERC-1271 signature mismatch.
If the facilitator violates \hyperlink{sr8}{\textbf{SR8}} by failing to ensure
that off-chain ERC-1271 validation is context-equivalent to the eventual
on-chain transaction, an attacker can induce a verification-settlement
divergence.
Since ERC-1271 validity is determined by arbitrary contract logic, 
a facilitator that relies only on off-chain \texttt{eth\_call} simulation may observe a successful verification even though the same proof fails when executed on chain during settlement.
Concretely, an attacker-controlled ERC-1271 validator contract can make \texttt{isValidSignature} depend on context-sensitive inputs that differ between off-chain calls and real transactions.
For instance, the ERC-1271 smart contract can require \texttt{block.basefee == 0} so that a misconfigured verifier or simulator reports success, yet settlement fails under real network base fee semantics.

The second root cause is ultimately manifested at the server's acceptance
boundary. We therefore distill two necessary server-side rules:
\begin{mybox}
\textbf{Server-SR1.}
  The server must return the protected resource to the client only after 
  settlement succeeds under its acceptance criteria.
\end{mybox}
\begin{mybox}
\textbf{Server-SR2.}
  If verification succeeds but settlement later fails, the server must roll back any 
  business logic effects triggered after verification and treat the request as unpaid.
\end{mybox}

\noindent\textbf{Asset theft attack.}
Asset theft is enabled by violations of \hyperlink{sr8}{\textbf{SR8}}.
If the facilitator fails to strictly validate and bind the server-declared
requirements to the corresponding proof fields, or accepts proofs whose on-chain
value-moving semantics are not explicit and unambiguous, an attacker can smuggle
unexpected value-moving operations into the sponsored settlement.
Concretely, this includes (1) inducing an attacker-directed transfer via a
malicious ERC-6492 contract signature (where settlement may execute additional
logic beyond a simple transfer), or (2) forcing the facilitator to pay rent/fees
for attacker-controlled accounts during an SPL Token transfer (e.g., ATA-related
side effects on Solana). 
\begin{figure*}[t]
  \centering
  \includegraphics[width=0.9\linewidth]{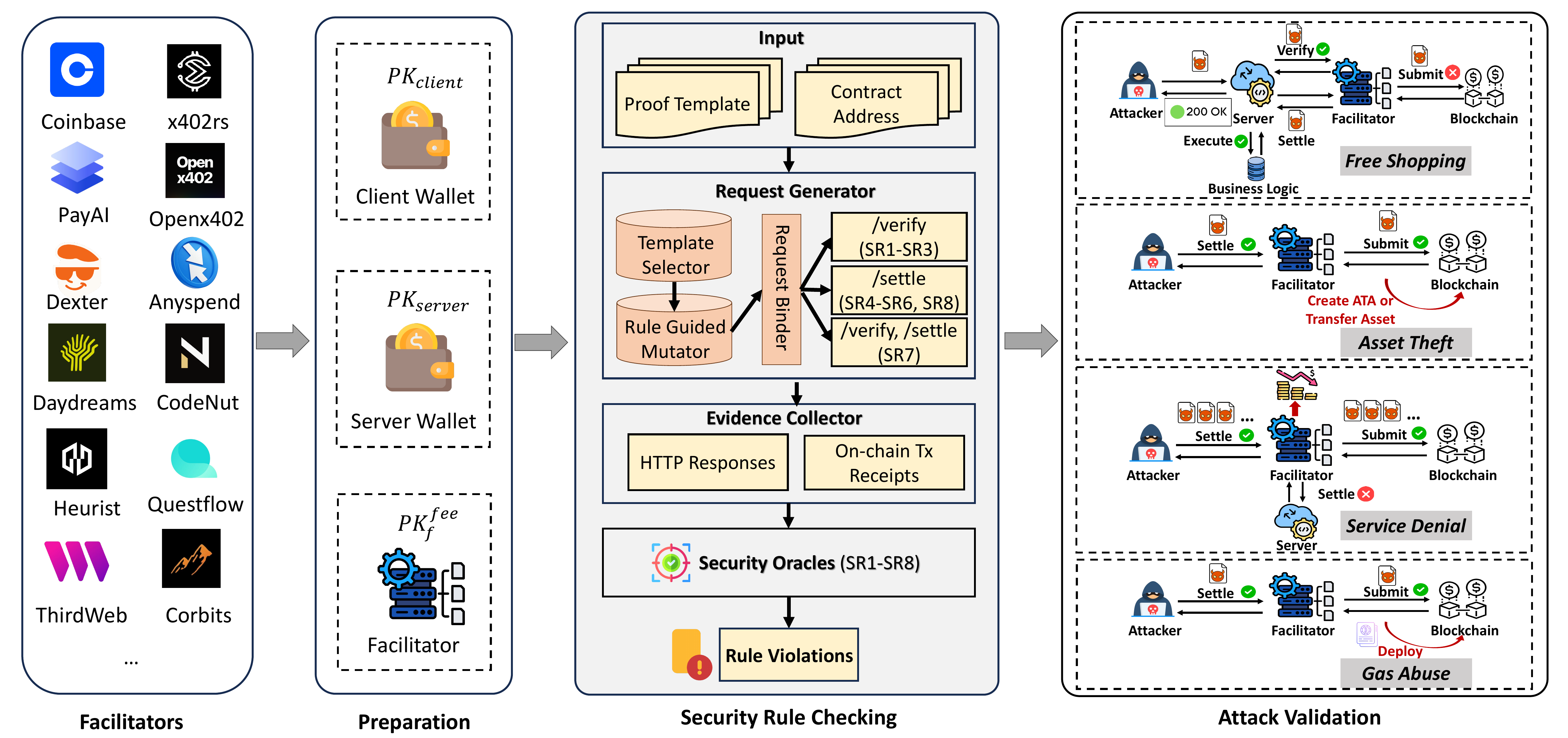} 
  \caption{Overview of \sysname{}.}
  \label{fig:systemoverview} 
\end{figure*}

\noindent\textbf{Service denial attack.} 
This attack targets availability by exhausting facilitator resources,
primarily by draining sponsor-paid gas/fees and inducing a high rate of failing
settlement submissions.
The attacker repeatedly submits proofs that are cheap to generate but expensive
for the facilitator to validate and/or submit on-chain.
We distill the following common enabling paths (each corresponding to a rule
gap that allows expensive failing submissions):
(1) \hyperlink{sr7}{\textbf{SR7}} combined with any gap in \hyperlink{sr1}{\textbf{SR1}}--\hyperlink{sr3}{\textbf{SR3}}. If the facilitator fails to reject invalid or non-fresh proofs at verification time under SR1 to SR3 or SR5, and also fails to perform re-validation at settlement time under SR7, it may sponsor settlement transactions that revert on-chain while still consuming gas;
(2) \hyperlink{sr5}{\textbf{SR5}} and \hyperlink{sr7}{\textbf{SR7}} gaps that fail to enforce SR5 fail fast checks and SR7 settle time re-validation. Under SR5, the facilitator should reject economically meaningless payments such as zero amount or dust, reject proofs that are not settleable such as insufficient balance, and enforce idempotency and a reasonable remaining validity bound to prevent replay and time window TOCTTOU. If these checks are permissive at \texttt{verify} and not re-checked at \texttt{settle} under SR7, an attacker can use repeated nonces, short validity windows, or balance dependent proofs to create verify pass but settle fail outcomes, and can also submit unbounded settlement requests that trigger sponsored execution with no meaningful value transfer;
(3) \hyperlink{sr8}{\textbf{SR8}} gaps via adversarial ERC-1271 logic that
induces repeated verify-settle inconsistencies and expensive failing
settlements;
(4) \hyperlink{sr8}{\textbf{SR8}} gaps via ERC-6492 signatures that trigger
additional deployment/initialization steps and amplify per-request cost or
failure rate;
and (5) \hyperlink{sr6}{\textbf{SR6}} gaps on Solana where fee-related parameters
(e.g., compute unit limit/price) or signing constraints are not fully validated,
allowing proofs that pass partial checks but fail or become prohibitively
expensive at submission time.




\noindent\textbf{Gas abuse attack.} 
Gas abuse exploits the facilitator's role as the fee sponsor by amplifying
sponsor-paid gas/fees per request beyond what is justified by the intended
payment. Unlike service denial (primarily availability disruption), gas abuse is
economically motivated: the attacker converts the sponsor’s budget into
attacker-chosen on-chain execution by steering \texttt{settle} into expensive
paths beyond a simple transfer (e.g., ERC-6492-style contract deployment).
This attack is enabled primarily by violations of \hyperlink{sr6}{\textbf{SR6}}
and \hyperlink{sr8}{\textbf{SR8}}: missing or insufficient gas/fee/compute parameter validation (\hyperlink{sr6}{\textbf{SR6}}), and insufficient
semantic binding between server-declared requirements and the actions executed
at settlement (\hyperlink{sr8}{\textbf{SR8}}).

\section{Methodology of \sysname{}}
\label{sec:method}

Guided by the security rules in~\autoref{sec:securityanalysis}, we propose
\sysname{}, a semi-automated framework for testing the authorization correctness and
execution safety of target facilitators.
\sysname{} follows a feature-aware, rule-guided workflow. It
first infers enabled features such as supported
networks and payment proof types, and then generates only applicable tests. As
illustrated in~\autoref{fig:systemoverview}, the workflow consists of
preparation (\autoref{sec:preparation}), systematic rule checking~(\autoref{sec:rulechecker}), and attack validation (\autoref{sec:attackvalidation}).


\subsection{Preparation}
\label{sec:preparation}
Our setup creates two wallets per network. The client wallet ($PK_{client}$)
generates payment proofs and must hold a small balance of the payment asset
(e.g., 1~USDC). The merchant wallet ($PK_{server}$) is used as the designated
payee in the payment requirement and as the intended on-chain recipient. Wallet
creation is a one-time step using standard key generation.

For each facilitator under test, we record its URL and, when required, obtain an
API key from its website and documentation. For Solana deployments, the client additionally
needs the facilitator fee payer public key ($PK_f^{fee}$) to construct proofs in
the required format. We obtain $PK_f^{fee}$ from the facilitator's \texttt{/supported}
endpoint.

\subsection{Security Rule Checker} 
\label{sec:rulechecker}
After setup, we test each facilitator for violations of the security rules in Section~\ref{sec:securityanalysis} across the x402 verification and settlement workflow.
To support custom features, heterogeneous networks, and proof variants, we implement \sysname{} as a suite of modular, rule-specific tests with tunable parameters.
As shown in~\autoref{fig:systemoverview}, \sysname{} takes payment-proof templates and testing contract addresses as inputs.
The templates follow the x402 protocol specification and cover the proof variants we test, including ERC-3009 \texttt{transferWithAuthorization} proofs, ERC-1271 smart-wallet signatures, and ERC-6492 wrappers for undeployed wallets.
Given these inputs, \sysname{} follows a six-step workflow.
First, \emph{capability discovery} infers the target deployment's supported networks, proof types, and optional features, and runs only applicable tests to reduce unnecessary queries and cost.
Second, \emph{template instantiation} creates x402 payment proofs from protocol-compliant templates, including ERC-3009 \texttt{transferWithAuthorization} proofs, ERC-1271 smart-wallet signatures, and ERC-6492 wrappers for undeployed wallets.
These templates are instantiated with deployment-specific parameters such as chain, asset, recipient, and amount.
Third, \emph{rule-guided mutation} modifies proof or requirement fields according to the security rule being tested.
Fourth, \emph{execution} sends the generated payloads to the facilitator's \texttt{/verify} and \texttt{/settle} endpoints.
Fifth, \emph{evidence collection} records HTTP responses and on-chain transaction receipts.
For example, \texttt{/verify} responses are expected to report whether the proof is valid, e.g., \texttt{"isValid": true/false}, while \texttt{/settle} responses are expected to report settlement status, e.g., \texttt{"success": true/false}, together with an on-chain transaction hash when settlement succeeds.
If a facilitator does not return a transaction identifier, we manually inspect the relevant chain activity from the configured client address to determine whether any settlement transaction was submitted.
Finally, \emph{security oracles} interpret the \texttt{/verify} and \texttt{/settle} outcomes for the rule-specific mutated payloads and classify each test outcome as pass, fail, or ambiguous.

The required manual effort is limited and controlled.
We manually derive proof templates from the x402 specification only once. \sysname{} then instantiates them automatically with deployment-specific parameters.
Most checks run automatically, and manual inspection is needed only for facilitator-specific JSON fields in their responses or for missing transaction identifiers.
The design is extensible: supporting a new network or proof format typically requires adding rule-specific test cases, not changing the checker architecture.
To improve efficiency, we often couple verification and settlement in the same
test run, allowing us to simultaneously infer feature support and check multiple
rules. 
To minimize ethical risks, we prioritize running tests on testnets whenever
possible. Testnet and mainnet deployments are configured identically, so our
testnet results carry over to mainnet behavior. 
We next describe each rule-checking module, including how \sysname{} constructs the mutated payload and how the observed \texttt{/verify}, \texttt{/settle}, and on-chain outcomes are mapped to rule violations.


\noindent\textbf{Network support tests.}
We perform live tests on Base and Solana, including both testnet and mainnet
deployments. For each network, we send a valid payment proof and check whether
both verification and settlement succeed. We consider the deployment supported
only if the facilitator returns a valid response and provides a transaction hash
that corresponds to a successful on-chain settlement.

\noindent\textbf{Validity window threshold tests (Base).}
This Base-specific check evaluates the validity window threshold enforced by target facilitators.
This threshold serves as a critical feature and security indicator: an overly small window implies that the facilitator accepts proofs likely to expire between verification and on-chain settlement, thereby triggering the Time-window TOCTTOU vulnerability and leading to free shopping or service denial (\autoref{sec:attack}).
To measure this, we construct ERC-3009 proofs with varying validity windows by setting \texttt{validBefore} to \texttt{now}+\(\Delta\), starting from \(\Delta=1\) second and increasing it. 
For each \(\Delta\), we invoke \texttt{verify} and \texttt{settle} and record whether the facilitator accepts the proof; the minimum accepted \(\Delta\) defines the facilitator's validity-window threshold.

\noindent\textbf{Payment field tampering tests (Base and Solana).}
Starting from a valid proof, we test two mutation modes: integrity mutations that change fields such as \texttt{scheme}, \texttt{network}, signature, or signer without re-signing, and binding mutations that change value-moving fields such as recipient or amount and then re-sign with \(PK_{client}\).
The mutated proof should be rejected by \texttt{/verify} and must not trigger on-chain submission.
We map outcomes by the violated property: accepting mutated payment requirements in \texttt{/verify} indicates an \hyperlink{sr1}{\textbf{SR1}} violation; accepting invalid signatures, wrong signers, or unauthentic authorization indicates an \hyperlink{sr2}{\textbf{SR2}} violation; reporting settlement success without a valid intended on-chain payment indicates an \hyperlink{sr4}{\textbf{SR4}} violation; submitting invalid or non-settleable proofs on chain indicates an \hyperlink{sr5}{\textbf{SR5}} violation; and submitting a proof after \texttt{/verify} rejects it indicates an \hyperlink{sr7}{\textbf{SR7}} violation.


\noindent\textbf{Zero amount settlement tests (Base and Solana).}
We run these tests on both Base and Solana by setting the settlement transfer amount to zero.
Such proofs should be rejected before any on-chain submission; submitting a zero-amount proof on chain indicates an \hyperlink{sr5}{\textbf{SR5}} violation.

\noindent\textbf{Balance and replay validity tests (Base and Solana).}
We test proofs that are syntactically valid but non-settleable due to insufficient balance or stale authorization.
For balance validity, we drain the client wallet and retry \texttt{/verify} and \texttt{/settle}; for replay/freshness validity, we test expired \texttt{validBefore}, invalid or far-future \texttt{validAfter}, and replayed nonces on Base, and stale \texttt{recentBlockhash} values or replayed transactions on Solana.
The facilitator should reject such proofs in \texttt{/verify} and refuse \texttt{/settle} before on-chain submission.
Accepting expired, premature, replayed, or stale authorization in \texttt{/verify} indicates \hyperlink{sr3}{\textbf{SR3}}; reporting settlement success without a valid intended payment indicates \hyperlink{sr4}{\textbf{SR4}}; submitting an insufficient-balance, expired, stale, or replayed proof on chain indicates \hyperlink{sr5}{\textbf{SR5}}; and submitting after failed verification or after the proof becomes stale indicates \hyperlink{sr7}{\textbf{SR7}}.

\noindent\textbf{Solana execution-safety tests.}
For Solana-supporting facilitators, we test instruction injection and sponsored-fee bounds.
For instruction injection, we add unexpected settlement instructions, such as ATA creation, extra instructions, or extra signers; acceptance indicates an \hyperlink{sr8}{\textbf{SR8}} violation, and unsafe on-chain submission or incorrect settlement success is additionally mapped to \hyperlink{sr5}{\textbf{SR5}} or \hyperlink{sr4}{\textbf{SR4}}.
For fee stress, we set excessive compute unit price or compute unit limit, e.g., \(10{,}000{,}000\); submitting the sponsored transaction indicates an \hyperlink{sr6}{\textbf{SR6}} violation because the sponsor fails to bound settlement cost.


\noindent\textbf{ERC-1271 and ERC-6492 contract-signature tests (Base).}
For Base/EVM facilitators, we first discover ERC-1271 and ERC-6492 support using baseline contract-signature proofs; a feature is supported only if \texttt{/verify} succeeds and \texttt{/settle} produces a successful on-chain settlement.
Given support, we run adversarial variants to test whether the facilitator accepts only explicit and unambiguous contract-signature semantics.
For ERC-1271, these include invalid contract signatures, off-chain/on-chain \texttt{isValidSignature} mismatches, and gas-heavy or unexpected validation logic.
For ERC-6492, these include gas-burning deployment or initialization logic, child-contract deployment, factory/calldata substitution, and unexpected transfer or execution semantics.
Accepting unauthentic contract authorization in \texttt{/verify} indicates an \hyperlink{sr2}{\textbf{SR2}} violation; reporting success without a valid intended on-chain payment indicates \hyperlink{sr4}{\textbf{SR4}}; sponsoring unbounded gas-heavy validation indicates \hyperlink{sr6}{\textbf{SR6}}; and accepting unexpected contract-signature execution semantics indicates \hyperlink{sr8}{\textbf{SR8}}.

\subsection{Attack Validation}
\label{sec:attackvalidation}

As shown in~\autoref{fig:systemoverview}, we illustrate the four new attacks with schematic attack flows.
Based on the test results and violation instances from~\autoref{sec:rulechecker}, we manually validate exploitability for each facilitator.
Because each rule can be violated in multiple ways, we do not infer attacks from SR violations alone; instead, we require attack-specific evidence.
\textbf{For asset theft,} we use two bounded evidence sources.
On Solana, we use an account without a pre-created ATA and inject an ATA-creation instruction into the settlement payload; if the facilitator submits the transaction and the ATA is created on chain, we treat it as evidence that attacker-supplied value-moving instructions can execute in the settlement context.
For ERC-6492 cases, we confirm whether the facilitator can be induced to submit a token-approval transaction on chain; a successful approval receipt indicates that value-moving authority can be granted through the settlement path.
We do not perform any subsequent transfer using the approval.
\textbf{For gas abuse,} we check whether the ERC-6492 settlement path executes attacker-controlled deployment logic.
Specifically, we inspect the on-chain receipt to determine whether the sub-contract is deployed or whether the gas-burning deployment/initialization path is executed, without attempting unbounded cost exhaustion.
\textbf{For free shopping,} we do not actively exploit third-party merchants for ethical reasons and therefore report high-risk evidence unless resource release is observed in an end-to-end merchant deployment.
In our controlled SDK setup, we determine potential resource release from the merchant HTTP response and inspect the server-side code for missing settlement-gated rollback.
Because real merchants may add application-specific checks, we conservatively treat SDK-level release-after-verify behavior combined with facilitator-side violations as high-risk evidence rather than a confirmed live exploit.
\textbf{For service denial,} we avoid active abuse for ethical reasons.
We therefore report high-risk evidence only when the facilitator accepts proofs that can verify but later fail, expire, or consume sponsored resources during settlement.

\begin{table}[t]
\centering
\caption{Overview of evaluated facilitators. We report the operator webpage along with transactions and volume from x402scan. * marks facilitators that also ship server SDKs.}
\label{tab:targets}
\resizebox{0.9\columnwidth}{!}{%
\begin{tabular}{@{}llll@{}}
\toprule
Facilitator & Webpage & Transactions & Volume \\ \midrule
Coinbase* & https://www.coinbase.com/ & 77.17M & \$26.85M \\
PayAI & https://facilitator.payai.network/ & 32.99M & \$4.58M \\
Dexter* & https://dexter.cash/ & 24.08M & \$4.62M \\
Daydreams & https://router.daydreams.systems/ & 11.82M & \$2.76M \\
Heurist & https://facilitator.heurist.xyz/ & 7.95M & \$30.04K \\
X402rs & https://github.com/x402-rs/x402-rs & 698.44K & \$1.50M \\
OpenX402 & https://open.x402.host/ & 697.35K & \$179.78K \\
Anyspend* & https://anyspend.com/x402 & 496.62K & \$100.07K \\
Codenut & https://www.codenut.ai/ & 477.92K & \$110.04K \\
Thirdweb* & https://thirdweb.com/ & 208.29K & \$116.16K \\
Corbits & https://corbits.dev/ & 153.62K & \$616.42 \\
Mogami* & https://facilitator.mogami.tech/ & 17.84K & \$305.26K \\
Ultravioleta Dao* & https://ultravioletadao.xyz/ & 4.76K & \$333.20 \\
xecho & https://www.xechoai.xyz/ & 4.38K & \$422.05 \\
Treasure & https://treasure.lol/ & 884 & \$248.70 \\ \bottomrule
\end{tabular}%
}
\end{table}

\begin{table*}[t]
\caption{
Security rule compliance and attack outcomes for evaluated x402 facilitators.
ERC-1271 and ERC-6492 denote support for the corresponding payer signature models on EVM (\fullcirc{} supported, \emptcirc{} not supported).
Valid Window Threshold reports the minimum remaining validity $T$ (in seconds) enforced at verification time, accepting a proof only if $(validBefore - now) \ge T$.
SR1 to SR8 indicate rule satisfaction (\greencheck{}) or violation (\redcross{}).
Attack columns summarize black-box results, where \exploitable{} denotes a directly validated exploit, \highrisk{} high-risk evidence without full exploitation, and \notexp{} not exploitable.
}
\centering
\label{tab:evalresults}
\resizebox{0.9\textwidth}{!}{%
\begin{tabular}{lccccccccccccccc}
\toprule
& \multicolumn{2}{c}{Sig Models} & \multicolumn{1}{c}{Validate Window} & \multicolumn{8}{c}{Security Rules} & \multicolumn{4}{c}{Attacks} \\
\cmidrule(lr){2-3}\cmidrule(lr){4-4}\cmidrule(lr){5-12}\cmidrule(lr){13-16}
Facilitator
& ERC-1271 & ERC-6492
& Threshold(s)
& SR1 & SR2 & SR3 & SR4 & SR5 & SR6 & SR7 & SR8
& Free Shopping & Asset Theft & Service Denial & Gas Abuse \\
\midrule
\rowcolor{gray!12}
1   &\fullcirc{}&\fullcirc{}& 5 & \greencheck{} & \greencheck{} & \greencheck{} & \greencheck{} & \greencheck{} & \redcross{} & \greencheck{} & \redcross{} & \highrisk{}  & \exploitable{} & \highrisk{}  & \exploitable{} \\
\rowcolor{gray!12}
2  &\fullcirc{} &\emptcirc{} & 6 &\greencheck{}&\greencheck{}& \greencheck{}  &  \redcross{} &  \redcross{}  & \redcross{}  & \redcross{} & \redcross{} & \highrisk{}   &   \notexp{}      & \highrisk{}  &  \notexp{}    \\
\rowcolor{gray!12}
3 &\emptcirc{}   & \emptcirc{} & 6 & \greencheck{} &\greencheck{}  &\greencheck{}  & \greencheck{} & \redcross{}  &  \redcross{}  & \redcross{} & \greencheck{}  & \notexp{}       &  \notexp{}      & \highrisk{}  &    \notexp{}   \\
\rowcolor{gray!12}
4 &\fullcirc{} &\emptcirc{} & 7 &\greencheck{}  &\greencheck{}  & \greencheck{}  & \greencheck{} & \redcross{}  &  \redcross{} & \redcross{}  & \redcross{}  & \highrisk{}   &     \notexp{}    & \highrisk{}  &   \notexp{}   \\
\rowcolor{gray!12}
5 &\fullcirc{} &\emptcirc{} & 7 & \greencheck{} &\greencheck{} & \redcross{}   &\greencheck{} & \redcross{}  & \redcross{} & \redcross{} & \redcross{}  & \highrisk{} &       \notexp{} & \highrisk{}  &  \notexp{}      \\
\rowcolor{gray!12}
6 &\fullcirc{}&\fullcirc{} & 7 & \greencheck{} &  \greencheck{}  &  \greencheck{}  &  \greencheck{} & \redcross{}  &  \redcross{} & \redcross{}  &  \redcross{} & \highrisk{}   &  \notexp{}     & \highrisk{}  & \exploitable{}   \\
7 &\fullcirc{} &\emptcirc{}& 7 & \greencheck{} &\greencheck{} & \greencheck{} & \greencheck{} & \redcross{}  & \greencheck{} & \redcross{} & \redcross{} & \highrisk{}   &  \notexp{}    & \highrisk{}  &    \notexp{}    \\
8 &\fullcirc{} &\emptcirc{} & 6 &\greencheck{}  &\greencheck{}  & \greencheck{}  & \greencheck{} & \redcross{}  &  \greencheck{} & \redcross{}  & \redcross{}  & \highrisk{}   &      \notexp{}   & \highrisk{}  &   \notexp{}   \\
9  & \fullcirc{}&\fullcirc{} & 6 & \greencheck{}  & \greencheck{}  & \greencheck{}  & \greencheck{}  & \redcross{}  &  \greencheck{}  & \redcross{}  &  \redcross{} & \highrisk{}    &    \notexp{}      & \highrisk{}  &   \exploitable{}   \\
10 & \emptcirc{} & \emptcirc{} & 7 & \greencheck{}  & \greencheck{}  & \greencheck{} &  \greencheck{} & \redcross{}  & \greencheck{}  &  \redcross{} &  \na{}   & \highrisk{}    &    \notexp{}  & \highrisk{}  &     \notexp{}    \\
11  &\emptcirc{}  &\emptcirc{} & 6 & \greencheck{}  & \greencheck{} & \greencheck{}  &\greencheck{} & \redcross{}  &  \greencheck{} & \redcross{} & \greencheck{}  &  \notexp{}    &       \notexp{}  & \highrisk{}  &    \notexp{}   \\
12  & \emptcirc{} & \emptcirc{} & 7 & \greencheck{}  & \greencheck{}  &\redcross{}  &    \greencheck{} & \redcross{}  &  \greencheck{} & \redcross{} & \na{}  & \highrisk{}       &     \notexp{}    & \highrisk{}  &     \notexp{}     \\
13 & \emptcirc{}  & \emptcirc{} & 3 & \redcross{}  & \redcross{} & \greencheck{}  &  \greencheck{}& \redcross{}  &\greencheck{}  &  \redcross{} & \na{}    & \exploitable{}    &     \notexp{}    & \highrisk{}  &   \notexp{}    \\
14 & \fullcirc{} & \emptcirc{} & 6 & \greencheck{}  & \greencheck{}  &  \redcross{}  & \greencheck{} & \redcross{}  &  \greencheck{}  & \redcross{}  & \redcross{} & \highrisk{}   &     \notexp{}     & \highrisk{}  &    \notexp{}    \\
15  &\emptcirc{}   &\emptcirc{}& 6 & \greencheck{} &\greencheck{} &\greencheck{} & \greencheck{} & \redcross{}  & \greencheck{}&\redcross{}& \na{} &  \notexp{}  &  \notexp{}       & \highrisk{}  &   \notexp{}   \\
\bottomrule
\end{tabular}%
}
\end{table*}

\section{Real World Evaluation}
\label{sec:realworldeval}
In this section, we present the first security study of \facilitatorNum{} real-world x402 facilitators used by over 60,000 sellers and 360,000 buyers.
We evaluate the major facilitators 
to study the following research questions: (1) Which security rules are violated? (Section~\ref{sec:rulesviolation}) (2) Do these violations introduce vulnerabilities that enable our proposed attacks? (Section~\ref{sec:newattack})

\noindent\textbf{Experimental settings.}
Using x402scan, a public ecosystem explorer for facilitator registration, we enumerated all publicly registered facilitator endpoints as of January 28, 2025, to ensure black-box testing coverage.
From this pool, we prioritized major deployments by ranking them according to x402-related transaction count and settlement volume during our measurement window.
We then removed two low-activity entries with fewer than 10 distinct buyers, and filtered out two facilitators with unreachable service URLs or those failing initial liveness probes.
The remaining \facilitatorNum{} facilitators form our evaluation set and is representative, accounting for 99\% of observed transactions and 98\% of total volume in the measurement window. 
\autoref{tab:targets} summarizes the targeted facilitators.
For each facilitator, we deployed a controlled merchant-side test service for free-shopping testing.
When a facilitator did not provide its own server SDK, we used the Coinbase x402 server SDK as the reference merchant-side implementation.

\subsection{Security Rule Violations}
\label{sec:rulesviolation}
Using \sysname{}, we detected \violateruleNum{} rule violations across the \facilitatorNum{} evaluated x402 facilitators.
To mitigate disclosure concerns, we report anonymized results in \autoref{tab:evalresults}, mapping each facilitator to a numeric ID.
The results show systematic non-compliance: every evaluated facilitator violates at least one security rule, and every rule is violated by at least one platform.
A rule may map to multiple concrete checkpoints (e.g., across networks and proof formats), so a single rule can yield multiple violation instances.
The most frequent failures involve \hyperlink{sr5}{\textbf{SR5}}, \hyperlink{sr7}{\textbf{SR7}}, and \hyperlink{sr8}{\textbf{SR8}}:
\sysname{} finds 14 deployments violating \hyperlink{sr5}{\textbf{SR5}} and \hyperlink{sr7}{\textbf{SR7}}, 
and 9 deployments violating \hyperlink{sr8}{\textbf{SR8}}.
For \hyperlink{sr5}{\textbf{SR5}} and \hyperlink{sr7}{\textbf{SR7}}, several deployments permit zero-value settlements or do not validate time/nonce strictly, enabling attacker-driven sponsor cost burn.
For \hyperlink{sr8}{\textbf{SR8}}, we observe weak freshness enforcement (e.g., overly permissive \texttt{validBefore} handling) that allows near-expired or invalid proofs to reach settlement, increasing failed-settlement rates and DoS-style cost amplification; in the worst case, invalid proofs accepted as paid can also enable free shopping or asset theft.
Importantly, violations are not limited to advanced checks: at least one deployment fails baseline requirements in \hyperlink{sr1}{\textbf{SR1}}--\hyperlink{sr4}{\textbf{SR4}}.

In addition, the evaluation is also efficient: the automated suite completes in under 10 minutes per target, with manual effort limited to initial configuration and evidence triage. Only four responses required manual adjudication due to custom response formats.
Because facilitators are black-box remote services and no ground-truth violation corpus exists, precision and recall are not well defined. We therefore report only evidence-backed violations using HTTP, facilitator, and, where applicable, on-chain evidence. False negatives may remain due to uncovered payment proofs (e.g., from Starknet), chain-specific or facilitator-specific behavior, merchant-specific integrations, and ethics-imposed limits on destructive testing. 
Thus, an SR pass means that the facilitator passed our implemented checks, not that it is generally secure.

\subsection{Discovered Attacks}
\label{sec:newattack}
After end-to-end attack validation and manual confirmation, we identified
\bugCount{} exploitable attack instances affecting \facilitatorNum{} facilitator deployments.
\autoref{tab:evalresults} summarizes four attack vectors discussed in~\autoref{sec:attack} and separates directly validated exploits (\exploitable{}) from high-risk evidence (\highrisk{}), where facilitator-side violations exist but ethical or deployment constraints prevent full live exploitation.
Asset theft and gas abuse can be validated through bounded ERC-6492 tests that expose value-moving or costly execution semantics.
For free shopping and service denial, we avoid harming third-party merchants or exhausting facilitator resources, and instead combine facilitator-side violations with local server-SDK experiments and attack-path analysis.
Ambiguous cases caused by unsupported proof formats,  third-party merchants logic, rate limits, or transient deployments are conservatively reported as high-risk rather than confirmed attacks.
Overall, sponsor-paid cost amplification and free shopping are the most prevalent, while asset theft is rarer but highest impact.

\noindent\textbf{Service denial.}
All evaluated facilitators exhibit a high risk of service denial and cost-amplification variants.
First, 14 out of 15 facilitators violate \hyperlink{sr5}{\textbf{SR5}} by allowing a malicious server to trigger economically meaningless settlements.
Second, nine facilitators support ERC-1271 signatures yet violate \hyperlink{sr8}{\textbf{SR8}}, allowing adversarial contract signatures to inflate verification and settlement costs (and potentially induce failures) via complex signature-validation logic.
Third, three facilitators support ERC-6492 signatures and violate \hyperlink{sr8}{\textbf{SR8}} by accepting proofs that can trigger additional on-chain deployment or initialization overhead (e.g., sub-contract creation).
On Solana, three facilitators violate \hyperlink{sr6}{\textbf{SR6}} by accepting multi-signer proofs without
bounding or fully validating the signer set (allowing an attacker to inject many co-signers) and by failing to constrain fee-related parameters (e.g., compute unit limit and price). These gaps can drive excessive fee burn and increase the likelihood of systematic submission failures.
Beyond these systemic issues, we observe concrete validation gaps.
Three facilitators, respectively, fail to validate the scheme, nonce freshness, and balance/account state, leading to violations of \hyperlink{sr1}{\textbf{SR1}}, \hyperlink{sr2}{\textbf{SR2}}, \hyperlink{sr3}{\textbf{SR3}},
\hyperlink{sr5}{\textbf{SR5}}, and \hyperlink{sr7}{\textbf{SR7}}.
These gaps allow attackers to craft proofs that repeatedly waste sponsor-paid gas via failing settlements.
We also find one facilitator violating \hyperlink{sr7}{\textbf{SR7}} by not enforcing the \texttt{validBefore} constraint immediately prior to settlement submission, leading to avoidable failures.
Finally, even when deployments satisfy the rules, many configure very short \texttt{validBefore} windows ($T \in [3,7]$ seconds).
Such short thresholds can still increase failure rates because \texttt{validBefore} is
checked at verification time, while settlement execution occurs later: after a proof passes verification, the facilitator must construct the settlement transaction, broadcast it, and wait for network propagation and inclusion.
This introduces a non-negligible delay between verification and on-chain execution. If the delay exceeds the
remaining validity margin enforced by the threshold, the transaction reaches execution with an expired
\texttt{validBefore} and fails, wasting sponsor-paid fees.

\noindent\textbf{Free shopping.}
We find widespread exposure to free shopping.
Among the evaluated facilitators, 10 cases are classified as high risks, 
and two cases are fully exploitable with end-to-end validation.
Both validated cases violate \hyperlink{sr4}{\textbf{SR4}}. 
In one, the facilitator fails to enforce client-side account balance constraints 
and returns valid for both verification and settlement.
In the other, an attacker can replay a previously successful payment proof, violating \hyperlink{sr4}{\textbf{SR4}} and \hyperlink{sr7}{\textbf{SR7}}. 
Notably, even if a replay is flagged as invalid during verification (e.g., due to a \texttt{validBefore} mismatch), the gap can still be exploited via concurrent verification request. 
We detail this in our case study~\autoref{sec:casestudy}.
The remaining 10 high-risk cases follow a \emph{verification-success / settlement-failure} pattern, where a proof is accepted at verification time but fails to settle on chain.
Whether this results in free shopping depends on merchant integration: after verification succeeds, the server may execute business logic and return the protected response without ensuring settlement succeeds or compensating for failed settlement.
We find that 9 facilitators violate \hyperlink{sr8}{\textbf{SR8}} by supporting the adversarial ERC-1271 signature, which can reliably induce this pattern (verification succeeds while settlement fails), and one facilitator exhibits verification flaws that violate \hyperlink{sr1}{\textbf{SR1}}
and \hyperlink{sr7}{\textbf{SR7}}.
Moreover, many deployments configure tight payment-validity thresholds, which increases the chance that a proof expires while the server is executing business logic. In this case, the proof can pass verification but become invalid by the time the settlement transaction is constructed, broadcast, and executed on-chain, leaving the server unpaid even when the facilitator's checks are otherwise correct.
In particular, we observe that the Coinbase Flask SDK (\(\le\) v0.2.1) continues request handling and returns the protected response immediately after verification succeeds, without gating the response on a successful settlement.
As a result, a verification-only acceptance path exists whenever settlement fails or expires.
Finally, none of the evaluated merchant SDKs implement an explicit rollback mechanism for business-logic side effects.
While x402 v2 introduces a callback interface, it is not specified as a rollback primitive, leaving safe compensation and atomicity to application logic.

\begin{table}[t]
\centering
\caption{Root-cause classification of reported findings.}
\label{tab:rootcause}
\resizebox{0.9\linewidth}{!}{
\begin{tabular}{l l}
\toprule
Finding & Root cause category \\
\midrule
SR1--SR4 violations & \begin{tabular}[c]{@{}l@{}}Facilitator implementation or deployment bugs \\ in proof binding, authentication and freshness\end{tabular} \\
SR5 violations & \begin{tabular}[c]{@{}l@{}}Implementation gaps at the facilitator-sponsored\\ settlement boundary\end{tabular} \\
SR6/SR8 violations & \begin{tabular}[c]{@{}l@{}}Chain/proof-specific execution semantics plus insufficient\\ implementation-side bounds and allowlists\end{tabular} \\
SR7 violations & \begin{tabular}[c]{@{}l@{}}Verify/settle split plus missing pre-settlement checks\end{tabular} \\
Free shopping & \begin{tabular}[c]{@{}l@{}} Merchant/SDK release-after-verify behavior,\\ often triggered by facilitator-side authorization failures\end{tabular} \\
Asset theft & \begin{tabular}[c]{@{}l@{}}Contract-signature ambiguity and \\ facilitators' settlement-semantics handling failures\end{tabular} \\
Service denial & \begin{tabular}[c]{@{}l@{}}Verify/settle timing and state divergence plus deployment \\ choices that expose availability impact\end{tabular} \\
Gas abuse & \begin{tabular}[c]{@{}l@{}}Facilitator-sponsored cost model plus insufficient\\ limits on attacker-influenced execution\end{tabular} \\
\bottomrule
\end{tabular}
}
\end{table}


\noindent\textbf{Gas abuse and asset theft.}
We identified three gas-abuse instances among facilitators that support ERC-6492 signatures. These facilitators violate \hyperlink{sr8}{\textbf{SR8}} by allowing counterfactual signatures to trigger sub-contract deployment during settlement.
We also found one asset theft instance stemming from the ERC-6492 handling.
Specifically, the facilitator accepts untrusted deployment metadata from the client and 
sends attacker-specified calldata to an attacker-specified address, turning settlement 
into an arbitrary-call primitive funded and signed by the facilitator.
More details are provided in our case study~\autoref{sec:casestudy}.
While our black-box tests did not observe Solana ATA-creation abuse, our on-chain measurement (\autoref{sec:furtheranalysis}) reveals suspicious patterns consistent with ATA-creation cost anomalies; several vendors quietly mitigated related issues after our disclosures.

Based on the above discussion, \autoref{tab:rootcause} distinguishes implementation bugs, risks from the verify/settle split or facilitator-sponsored settlement model, and deployment-dependent integration choices. SR1--SR4 mainly reflect facilitator implementation or integration bugs, showing that authorization correctness is fragile in the wild. SR5--SR8 capture settlement and execution-safety risks caused by chain- or proof-specific semantics, the verify/settle split, and facilitator-sponsored settlement. At the attack level, free shopping depends on release-after-verify patterns without settlement-gated rollback; asset theft is primarily an implementation failure amplified by underspecified contract-signature settlement semantics; service denial arises when proofs verify but later fail, expire, or consume resources during settlement; and gas abuse reflects sponsored-cost ambiguity with insufficient bounds. Overall, x402 spans multiple parties, layers, proof formats, and signature models, creating a large validation surface that makes end-to-end atomicity 
and cost bounding difficult in practice.





\subsection{Case Study}
\label{sec:casestudy}
This section presents two representative case studies showing how implementation gaps translate into end-to-end exploits in real-world x402 deployments.

\noindent\textbf{Free shopping.}
We identified a validated free-shopping vulnerability caused by inconsistent freshness and replay enforcement across a facilitator's \texttt{/verify} and \texttt{/settle} endpoints.
The vulnerability arises because \texttt{/verify} is stateless and does not reserve the payment nonce, while \texttt{/settle} is not atomically bound to a unique verification result.
An attacker can therefore reuse the same valid payment proof in many concurrent resource requests before the nonce is consumed.
These requests may all pass verification, and repeated settlement attempts may be accepted or interpreted by the merchant as successful authorization.
Consequently, the merchant can execute business logic and return protected resources multiple times even though the buyer authorized only one payment.
This behavior violates \hyperlink{sr7}{\textbf{SR7}} because the same proof is not enforced as one-time-use, and violates \hyperlink{sr4}{\textbf{SR4}} because settlement success is not uniquely tied to a fresh, intended on-chain payment.
The resulting impact is end-to-end free shopping: repeated service fulfillment can be obtained under a single payment authorization, including across distinct requests and potentially distinct resources.
We disclosed the issue, and the vendor is working on a fix.

\noindent\textbf{Asset theft.}
We identified an ERC-6492-related asset theft vulnerability in a top-volume x402 facilitator's EVM implementation.
The root cause is the blind trust of the attacker-supplied ERC-6492 deployment.
During verification, the facilitator does not cryptographically validate the ERC-6492 wrapper or inner signature. 
During settlement, however, it trusts attacker-supplied deployment metadata (e.g., \texttt{factoryAddress} and \texttt{factoryCalldata}) and submits \texttt{(to=factoryAddress, data=factoryCalldata)} on the blockchain. 
Because \texttt{factoryAddress} is attacker-controlled and need not be a legitimate wallet factory, the attacker can repurpose these fields to encode an arbitrary contract call, such as token \texttt{approve} or \texttt{transfer}.
This turns settlement into an attacker-chosen transaction executed by the facilitator's settlement EOA.
Consequently, the attacker can directly move assets held by the facilitator EOA or create persistent token approvals that can be drained later.
Moreover, the API may report settlement failure even when the attacker-specified transaction has already been broadcast and confirmed on chain, so harmful side effects can occur despite a failure response.
We validated the issue with minimal proof-of-concept transactions under our control, did not exploit it beyond confirming impact, and disclosed it to the vendor.
The vendor acknowledged the issue and is deploying mitigations to tighten ERC-6492 verification and constrain settlement semantics.

\section{Ecosystem Risks Measurement}
\label{sec:furtheranalysis}
We present an empirical measurement study of mainnet x402 transactions 
on Base and Solana to characterize ecosystem trends and surface on-chain signals of risks.
Section~\ref{sec:realworldeval} reports validated facilitator-side findings for evaluated platforms, while this section provides complementary ecosystem-level context from address-based mainnet data.
Importantly, our measurement is not intended to label historical transactions as confirmed attacks.
Instead, it provides ecosystem-level risk evidence for current facilitator-mediated x402 deployments.
We aim to answer the following research questions.

\textbf{RQ1:} How large is x402 usage on mainnet and how concentrated are facilitators?

\textbf{RQ2:} How often does settlement fail and what sponsor-paid costs does it impose?

\textbf{RQ3:} What on-chain patterns are consistent with x402-specific risks?

\textbf{Dataset.}
We collect x402-related on-chain transactions from Base and Solana starting in May 2025, coinciding with the x402 proposal (Base block height $\approx$ 30M).
On Base, we analyze a contiguous window of 10M blocks (30M to 40M) and extract the corresponding Solana transactions over the same time period.
To identify relevant traffic, we compile a list of all facilitator addresses registered on x402scan.
Since facilitators typically register on this public explorer to ensure discoverability, we use interaction with a registered facilitator as our primary inclusion criterion.
This approach yields a total of 91{,}507{,}800 transactions on Base and 28{,}171{,}386 on Solana.
Unless otherwise specified, we normalize all ETH- and SOL-denominated monetary values using spot prices as of Jan.~1,~2026 (ETH $\approx$ \$3,000, SOL $\approx$ \$125)~\cite{coingecko}.

\textbf{Attribution scope and limits.}
We identify on-chain x402 transactions using facilitator-address matching together with x402-relevant settlement filters, such as ERC-3009 function selectors (e.g., \texttt{0xcf092995}) and payment-field parsing. 
This methodology provides broad coverage of facilitator-mediated x402 activity, but it does not provide ground-truth labels for individual transactions.
Rare false positives may still arise from mixed facilitator workloads, testing or maintenance transactions, or non-x402 ERC-3009 and SPL-token interactions involving the same addresses.
Although no ground truth is available to quantify this effect precisely, we expect false positives to be limited because inclusion requires both facilitator-address matching and x402-relevant settlement or payment-structure evidence.
Our analysis targets publicly discoverable facilitator-mediated deployments. 
Thus, it may miss unregistered facilitators, proxy contracts, address rotation, or newly deployed addresses absent from our collection snapshot.
Our unique server counts are likewise address-level operational measures based on parsed payee or server addresses.

\subsection{Ecosystem Centralization (RQ1)}
\begin{figure}
  \centering
  \includegraphics[width=1\linewidth]{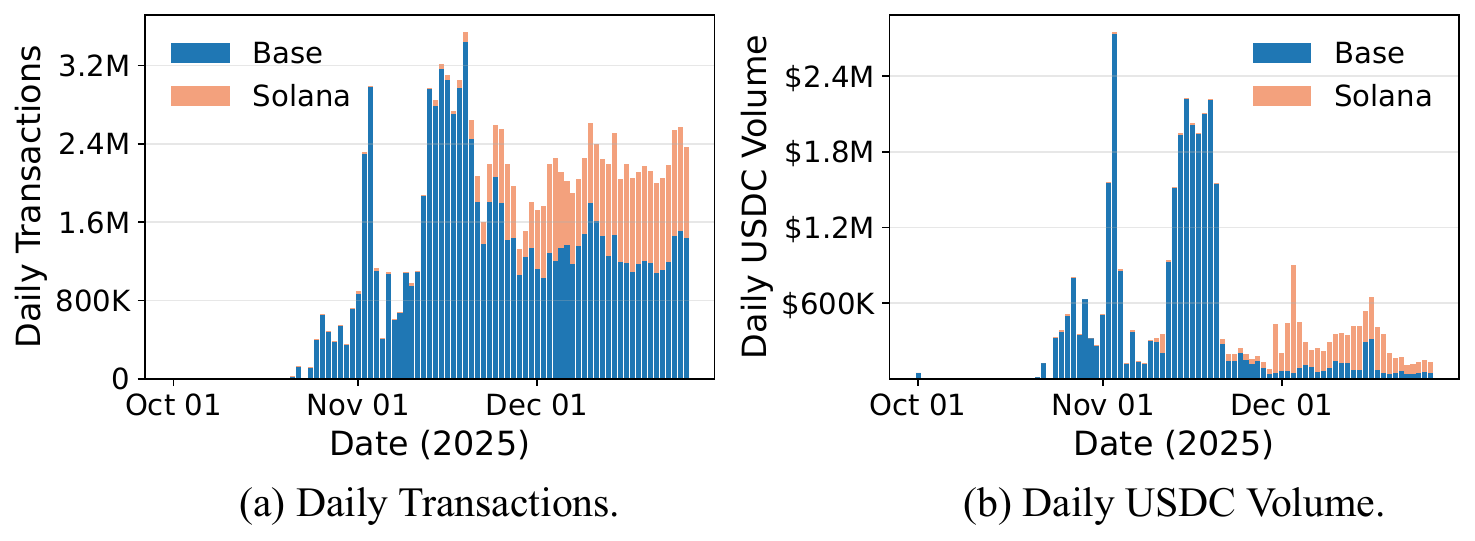} 
  \caption{Growth of x402 transaction activity and USDC payment volume across Base and Solana (Oct.~01--Dec.~26, 2025).}
  \label{fig:trend_of_transactions} 
\end{figure}

Although x402 transactions first appear on chain as early as May 2025, activity remains sparse until an inflection in early October, after which usage increases rapidly and stays active on both Base and Solana through Oct.~01 to Dec.~26, 2025. Figure~\ref{fig:trend_of_transactions} shows that transaction counts and USDC payment volume grow quickly from mid-October and remain high thereafter, peaking at about 3.5M transactions per day and over \$2.7M daily USDC volume.
Adoption starts on Base, but Solana surpasses Base in transaction count from late November onward, while Base continues to contribute substantial payment value. This indicates that x402 remains highly active on both networks with different usage profiles.

\begin{figure}
  \centering
  \includegraphics[width=1\linewidth]{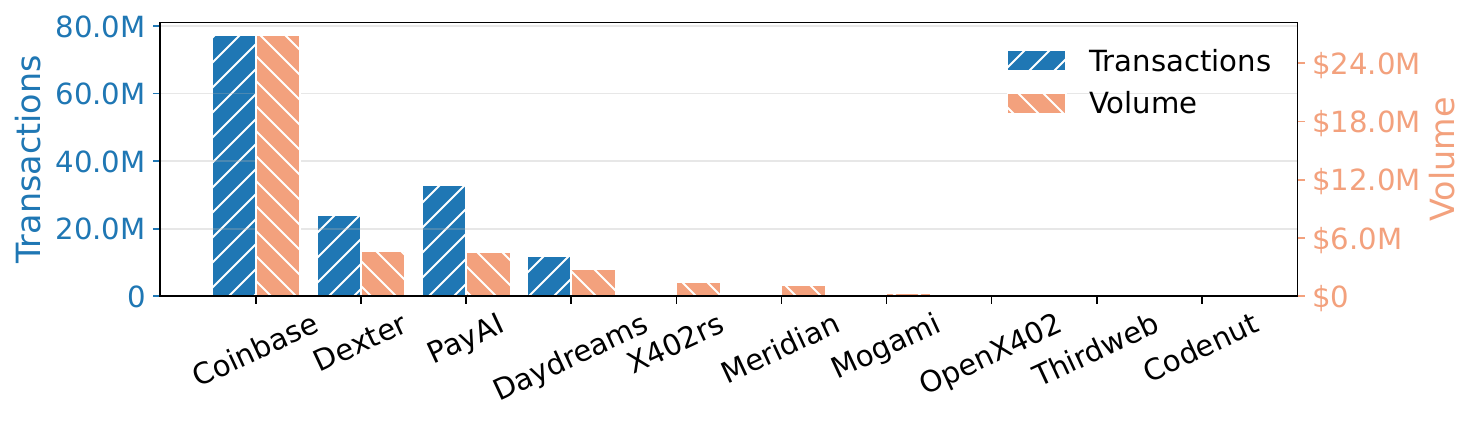} 
  \caption{Top-10 x402 facilitators by transaction count and payment volume.}
  \label{fig:top10_facilitators} 
\end{figure}

Figure~\ref{fig:top10_facilitators} presents the top-10 x402 facilitators ranked by transaction count and payment volume.
The ecosystem exhibits pronounced facilitator-level concentration, with Coinbase dominating both dimensions by a large margin, accounting for $77.17$M transactions and \$26.85M in payment volume. A small number of secondary facilitators (e.g., PayAI, Dexter, and Daydreams) form a distant second tier, while the remaining participants constitute a long tail with orders-of-magnitude lower activity.

Beyond facilitator-level concentration, we further examine whether servers connect to multiple facilitators. Across $53{,}576$ unique servers observed in our dataset, only $3{,}629$ ($6.77\%$) are associated with more than one facilitator, yielding an average of $1.089$ facilitators per server. This indicates that over $93\%$ of servers are exclusively bound to a single facilitator, revealing strong structural lock-in on the supply side. Together, these findings reveal ecosystem-level fragility. Concentrated facilitator market share and limited server multihoming mean that flaws in a leading facilitator translate into systemic exposure, affecting numbers of servers and clients.

\subsection{Settlement Failures and Costs (RQ2)}

\begin{figure}
  \centering
  \includegraphics[width=1\linewidth]{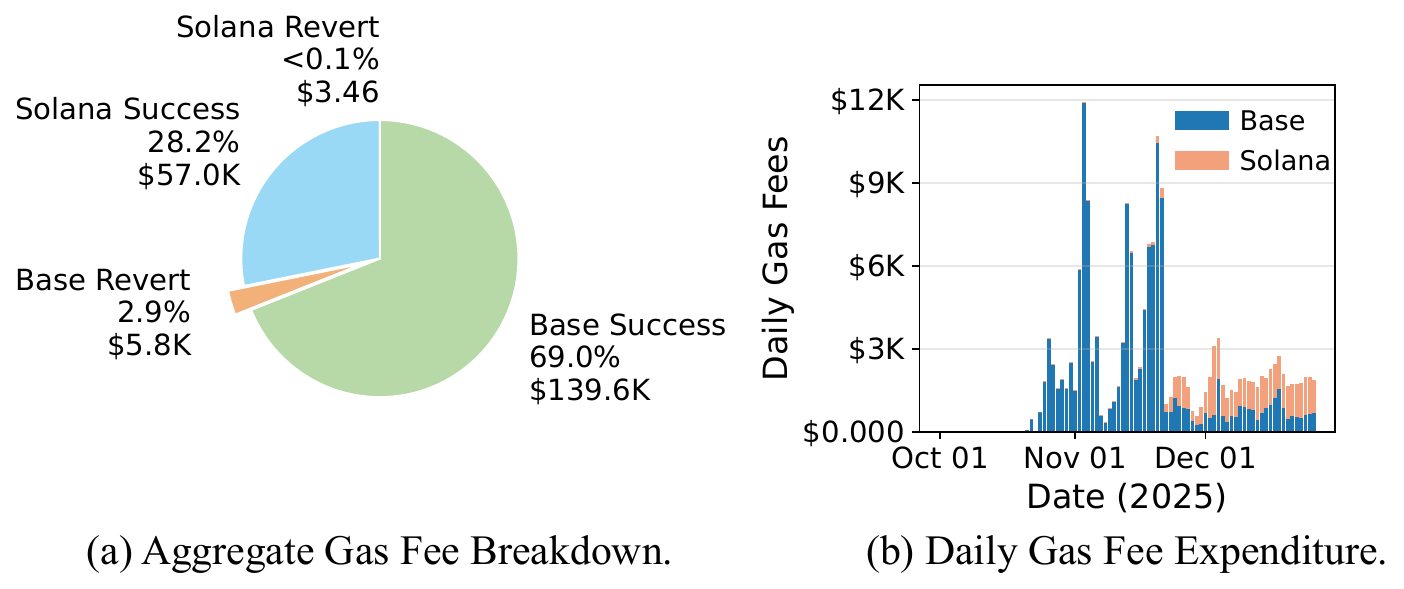} 
  \caption{Gas consumption of x402 across Base and Solana (Oct.~01--Dec.~26, 2025).}
  \label{fig:gas_consumption} 
\end{figure}

We examine settlement failures and their costs. Because facilitators sponsor on chain fees, reverted settlements directly translate into unrecoverable sponsor paid loss. As shown in Figure~\ref{fig:gas_consumption}, x402 has already incurred about \$202K in on chain fees, with Base contributing \$145.4K (71.9\%) and Solana \$57.0K (28.2\%). As shown in Table~\ref{tab:x402_fail_rate}, although reverts are a small fraction of activity (1.99\% on Base and $<0.1\%$ on Solana), they still burn thousands of dollars in gas and fees, making failures economically consequential in practice.

\begin{table}
\centering
\setlength{\tabcolsep}{8pt}
\caption{Revert statistics of x402 transactions on Base and Solana (Oct.~01--Dec.~26, 2025).}
\label{tab:x402_fail_rate}
\resizebox{0.36\textwidth}{!}{%
\begin{tabular}{lrrr}
\toprule
Network & Success & Reverted & Revert Rate
 \\
\midrule
Base   & 91{,}507{,}800 & 1{,}857{,}949 & 1.99\% \\
Solana & 28{,}172{,}878 & 5{,}148 & 0.018\% \\
\bottomrule
\end{tabular}
}
\end{table}

\begin{figure}
  \centering
  \includegraphics[width=0.85\linewidth]{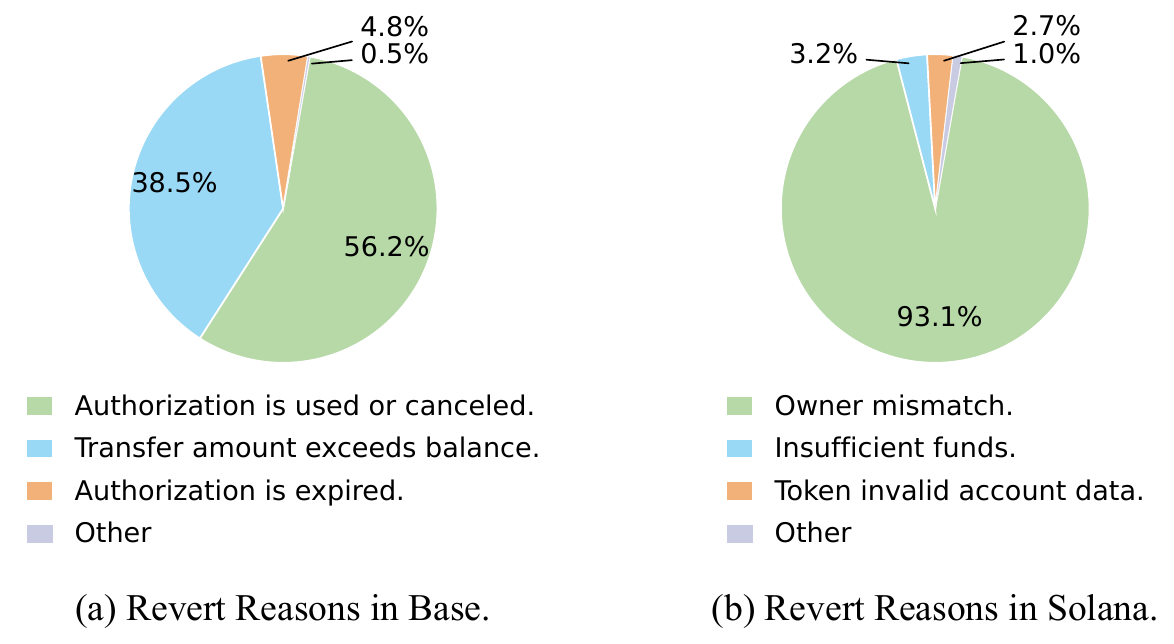} 
  \caption{Revert reason distribution of x402 transactions on Base and Solana (Oct.~01--Dec.~26, 2025).}
  \label{fig:revert_reason_pies} 
\end{figure}

We therefore further break down revert causes to understand the dominant failure modes and their security implications.
Figure~\ref{fig:revert_reason_pies} breaks down the main revert reasons on each network. On Base, failures are dominated by application-level payment logic, with authorization being used or canceled accounting for 56.2\% of reverts, followed by ERC20 transfers exceeding balance (38.5\%), while authorization expiration contributes an additional 4.8\%. In contrast, Solana exhibits a sharply concentrated failure profile, where 93.1\% of reverts are caused by owner mismatch, with the remainder mainly due to insufficient funds and invalid account data. 
For the most prevalent Base failure, \texttt{authorization is used or canceled}, 1{,}047{,}753 reverts originate from only 57 senders (top 5: 57.07\%, top 10: 76.41\%).
Although the address-based data alone cannot rule out benign operational causes such as retries or batch processing, this extreme skew is consistent with automated replay by a small set of entities.

Observability varies across attacks: free shopping is mostly off-chain, ERC-6492-style asset theft and gas abuse are only partially visible on-chain (often noisy due to heterogeneous benign deployment paths and off-chain failures), and service denial leaves little on-chain evidence.
In contrast, ATA-creation risks on Solana leave observable signals through \texttt{create\_associated\_token\_account} related instructions in settlement transactions. We therefore analyze anomalous ATA creation patterns. 
 ATA initialization requires a fixed rent-exempt deposit (0.00203928~SOL, $\approx$ \$0.25)~\cite{SolDoc}, which early x402 workflows shift from end users to facilitators. This subsidy can enable near-zero cost mass ATA creation for clients while externalizing locked capital to facilitators, and deposits can later be reclaimed via \texttt{CloseAccount}, enabling repeated create and close churn.
Table~\ref{tab:ata_rent} lists 37{,}959 ATA creation events, corresponding to approximately \$9{,}489.75 in cumulative deposits locked by facilitators. 
Daydreams, PayAI, and Dexter collectively account for over 80\% of all ATA creations. Although each individual ATA requires only a modest deposit, the aggregate locked capital becomes non-trivial at scale. This concentration makes a few facilitators the main points of exposure for sponsor-paid ATA rent abuse risk and cost amplification, since they cover most ATA rent deposits.


\subsection{On-Chain Risk Signals (RQ3)}
\begin{table}
\centering
\small
\caption{ATA rent events by facilitator. Each ATA creation incurs a fixed rent-exempt cost of 0.00203928~SOL. USD values are estimated using \$125/SOL.}
\resizebox{0.38\textwidth}{!}{%
\begin{tabular}{lrrr}
\toprule
Facilitator & ATA Creations & Rent (SOL) & Rent (USD) \\
\midrule
Daydreams        & 17{,}041 & 34.75 & \$4{,}344 \\
PayAI            & 10{,}901 & 22.23 & \$2{,}779 \\
Dexter           & 7{,}959  & 16.23 & \$2{,}029 \\
UltravioletDAO   & 1{,}518  & 3.10  & \$388 \\
AurraCloud       & 537      & 1.10  & \$138 \\
Anyspend         & 3        & 0.006 & \$0.75 \\
\bottomrule
\end{tabular}
}
\label{tab:ata_rent}
\end{table}

\begin{figure}
  \centering
  \includegraphics[width=0.65\linewidth]{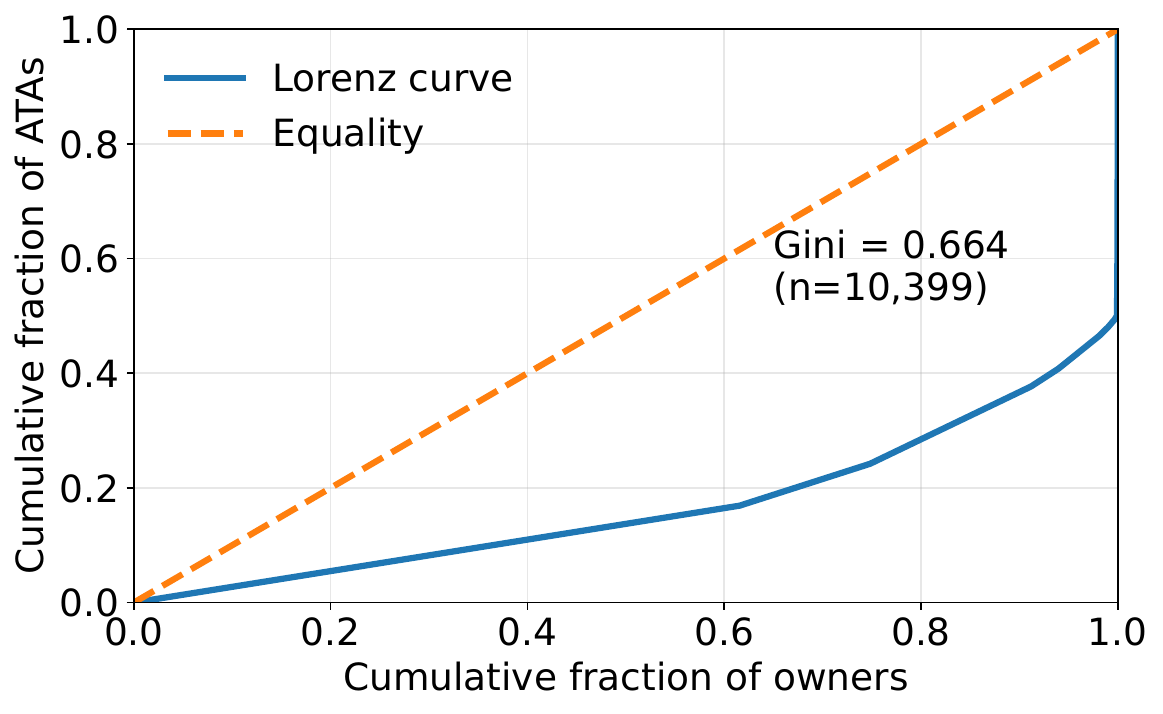} 
  \caption{Concentration of ATA creations across owners.}
  \label{fig:ata_lorenz_gini} 
\end{figure}

We next examine ATA creations across owners. Table~\ref{tab:ata_ranges} reveals a heavy-tailed pattern: most owners create fewer ATAs, but five entities each create more than 1,000. Moreover, 60.59\% of all ATAs are closed after fewer than three token transfers, indicating create and abandon churn rather than sustained usage. Figure~\ref{fig:ata_lorenz_gini} further quantifies this concentration with a Gini coefficient of 0.664. Together, the extreme tail behavior and rapid closure patterns show that a small minority drives a disproportionate share of ATA creation, which is consistent with automated abuse of subsidized account creation.

\begin{table}
\centering
\small
\caption{Distribution of ATA creations per owner. Owners are grouped by the number of ATAs they created to highlight extreme tail behavior.}
\resizebox{0.38\textwidth}{!}{%
\begin{tabular}{p{1.3cm}rrrrr}
\toprule
ATAs/Owner
 & 1 & 2--10 & 11--100 & 101--1000 & $>1000$ \\
\midrule
Owners & 6,387 & 4,002 & 1 & 4 & 5 \\
\bottomrule
\end{tabular}
}
\label{tab:ata_ranges}
\end{table}

\section{Discussion}
\label{sec:discussion}
\noindent\textbf{Lessons and mitigation.}
Our study focuses on facilitators' authorization correctness and execution safety. 
x402 is fast-evolving, leaving specification gaps and implementation drift across proof formats, signature models, and merchant--facilitator integrations.
Our security rules and evaluation suggest three practical mitigation directions.
(1) Bind verification to settlement, e.g., through nonce reservation, short-lived server-bound tokens, and re-checking time/state constraints before settlement.
(2) Treat client-provided fields as hostile execution inputs, and enforce strict validators, allowlists, and constrained transaction shapes, especially for ERC-1271/6492 paths.
(3) Bound sponsor-paid costs by default, by capping fee parameters, rejecting non-settleable or economically meaningless payments, and preflight-checking settlements that are likely to fail. Servers should gate fulfillment on settlement success or implement explicit rollback/compensation on failure.
Beyond EVM and Solana, x402 specifications and SDKs should make the trust boundary among verification, business execution, and settlement explicit, and provide default-safe templates for proof validation, settlement reporting, and sponsored execution. 
Our rules can be instantiated for other settlement environments by adapting them to network-specific proof formats, fee models, and execution constraints, providing a practical baseline for hardening x402 deployments.

\noindent\textbf{Gas-abuse impact.}
The impact of gas abuse depends on deployment-specific settlement economics, as facilitator pricing and reimbursement policies are heterogeneous, ranging from usage-based pricing to to fee-free or gas-sponsored settlement. If facilitators sponsor fees without reliable
reconciliation or chargeback, attacker-induced settlement can become direct sponsor loss. If costs are later reconciled with servers, the same behavior is viewed as cost shifting, liquidity pressure, or denial of service. The adversary role also matters. Malicious clients mainly abuse server-facing payment flows, whereas malicious or impersonated servers can steer facilitator-sponsored execution more directly.

\noindent\textbf{Scope and trust assumptions.} We assume facilitators as trust-bearing intermediaries.
A malicious or colluding facilitator represents a trust-failure model that is distinct from the protocol non-compliance problem studied in this paper.
In x402, the facilitator is not a custodian: the buyer signs the payment payload and settlement occurs on chain, so the facilitator cannot directly move buyer funds beyond the signed authorization or settle tampered payment fields as valid. A malicious facilitator could still cause service-level integrity or availability failures, e.g., by misreporting payment status, rejecting valid payments, or delaying settlement. Such risks are better handled through deployment controls, including reputable facilitator selection, on-chain settlement cross-checking, verifiable logs, transaction limits, fallback facilitators, or self-hosting.

\noindent\textbf{Future work.}
We will extend our black-box tool along three directions. First, we will broaden coverage to additional facilitators and track evolving SDK versions and default configurations to detect regressions over time. Second, we will expand network support to new settlement environments such as Starknet by instantiating our rules with network-specific payment proofs, fee models, and execution constraints. Third, we are already engaging with Coinbase and other ecosystem stakeholders to integrate our rule checks into development and pre-deployment validation workflows, and to distill the lessons into reusable best-practice guidance and reference specifications for servers, clients, and facilitators. 
Fourth, we will explore LLM-assisted testing for PoC generation, and response summarization to improve automation, while keeping all live tests human-reviewed and rate-limited.

\section{Related Work}
\label{sec:relatedwork}


\noindent\textbf{Payments system security.}
Prior work studies payment security in both card-present and online settings.
For card payments, research on EMV analyzes protocol features and extensions and
associated requirements (e.g., PCI DSS), and proposes defenses such as
privacy-preserving protocols and relay-attack mitigations~\cite{pavlides2025more,rahaman2019security,bursuc2023provably,olivier2025pays,boureanu2020security,coppola2024pure}.
For online payments in mobile apps and Web services, studies examine
application-layer APIs and workflows, including formal analyses of the W3C Web
Payment APIs and logic flaws in payment aggregation services, as well as
ecosystem-scale risks and mitigations~\cite{do2022formal,chen2019devils,kumar2020security,lou2021messy,abi2021doing}.
In contrast, our work focuses on application-layer payments for Web APIs and
autonomous agents, and characterizes the security of facilitator-mediated
verification and on-chain settlement in x402 deployments.

\noindent\textbf{Smart contract and on-chain security}
One major direction in blockchain security research is contract-centric and on-chain vulnerability analysis.
On EVM, work evaluates vulnerability scanners and
develops learning- and execution-based techniques, characterizes vulnerability
classes and deployment patterns, and proposes automated patching
approaches~\cite{sendner2024large,sendner2023smarter,wang2024smartinv,li2024enhancing,wang2023automated,babel2023lanturn,sun2024all,zhang2024demystifying,pan2025collisionrepair}.
Beyond EVM, studies fuzz Solana contracts and investigate resource-model DoS
weaknesses, while interoperability and atomic-swap work highlights cross-chain
settlement complexity~\cite{smolka2023fuzz,luo2024towards,augusto2024sok,thyagarajan2022universal}.
In contrast, we study x402 as an application-layer payment protocol whose blockchain settlement introduces security requirements largely orthogonal to contract-level bugs.
In addition, although some underlying primitives (e.g., ATA creation) are well known, we show that under the x402 workflow they induce new, exploitable failure modes due to weak verification-settlement binding and sponsored settlement semantics.
\section{Conclusion}
\label{sec:conclusion}
x402 is rapidly gaining real-world adoption, yet the security posture of facilitators remains poorly understood. 
This is especially concerning because facilitators sit on the critical path 
across many merchants. 
As a result, buggy or inconsistent facilitators
can misauthorize access or trigger unsafe settlement outcomes at scale.
To address this gap, we distill a checkable set of security rules for payment
correctness and execution safety and evaluate \facilitatorNum{} major facilitators, uncovering \bugCount{} vulnerabilities. We find that
non-standardized semantics of being ``paid'' and ``safe to settle'' lead to
check--execute mismatches that manifest as four attack classes.
Our key takeaway is that 
hardening
x402 requires a shared, rule-based security baseline across networks, proof
formats, and integration patterns.




\section{Acknowledgements}
\label{sec:acknowledgements}

We thank our reviewers and shepherd for their insightful feedback.
This project was supported, in part, by the New Generation Artificial Intelligence-National Science and Technology Major Project under No. 2025ZD0123503, SNSF 200021-236559 (LinSpecteur), PCEGP2\_186974, NSFC under No. U2441239 and U24A20336.
\appendix
\section*{Ethical Considerations}
We carefully considered the ethical implications of this research. We identify key stakeholders and impacts, describe our safeguards, and explain our decision to publish.

\noindent\textbf{Stakeholders and Potential Impact.}
Our study focuses on systematic security risks in x402 shared infrastructure and practical guidelines for facilitator and server implementations. Relevant stakeholders include facilitator operators, SDK maintainers, server/client developers, end merchants/customers, and blockchain network
participants.
(1) Our study most directly affects \emph{facilitator operators}, because we performed controlled tests against their live endpoints and reported issues that may carry limited operational burden and reputational risk. We mitigate
this impact through bounded, rate-limited, and testnet-first validation when available, and through responsible disclosure.
(2) For \emph{open-source SDK maintainers}, our interaction is non-invasive: we deployed SDKs locally to reproduce workflows and assess integration risks, with the goal of providing concrete guidance and test criteria.
(3) For \emph{server/client developers}, we did not interact with production systems. The expected impact is downstream security improvement through guidance on safe verification and settlement handling.
(4) For \emph{end merchants and customers}, we had no direct interaction, no access to user data, and did not attempt to obtain protected resources. Our facilitator tests used only our own accounts and resources.
(5) For \emph{blockchain network participants}, including validators and RPC providers, our controlled validation introduced only a small number of additional on-chain transactions. We also collected public on-chain data for
measurement, but did not access or infer non-public user information, such as off-chain identities, account credentials, or protected service content. Because our interactions were bounded and rate-limited, we do not expect
measurable impact on network health or RPC availability.

\noindent\textbf{Responsible Disclosure.}
When we found security issues in facilitators or server-side integrations, we followed responsible disclosure.
In January 2026, we privately reported our findings to 14 of the 15 affected parties, with enough technical detail to reproduce and validate each issue, and we coordinated mitigation timelines.
One evaluated facilitator xEcho did not provide a practical security reporting channel (e.g., no security contact or disclosure process), so we were unable to file a confidential report through normal means.
As of February 6, 2026, Coinbase, PayAI, and Mogami have acknowledged our reports and confirmed six distinct vulnerabilities, and they have already fixed some of the issues (with others still being addressed). Since these three vendors' services/SDKs are used by many downstream servers and clients in the x402 ecosystem, their fixes can benefit a broad set of deployments.
We are not releasing exploit-enabling details for unpatched issues, including target-specific payloads or reproduction steps that would materially facilitate abuse, and we will continue working with the remaining vendors as remediation progresses.


\noindent\textbf{Experiments with Live Systems.}
Our live-system experiments were intentionally conservative. We primarily
relied on public on-chain data and interacted with deployed endpoints only
when needed to validate suspected rule violations. When available, we first
used our own open-source deployments or testnets. We followed four
guardrails.
\emph{(1) Own identities and resources.}
For facilitator-facing tests, we acted as both server and client using
only accounts we control. All requests, proofs, and settlement transactions
were scoped to our own accounts and resources. We did not access, or
attempt to access, non-public user data or third-party paid resources.
\emph{(2) Rate limits and minimal probing.}
All active tests were rate-limited to one request every two seconds, and
each checker used the minimum interactions needed to collect evidence. Most
rule checks were repeated twice, three times only when initial results were
inconsistent, and five times for time-sensitive validity-window and
TOCTTOU checks. The only exception was one concurrency-dependent PoC, for
which we issued a single burst of 10 concurrent requests against the only
applicable target. We would have stopped testing upon signs of instability,
abnormal resource consumption, or unintended side effects, and observed no
service disruption.
\emph{(3) Testnet-first validation.}
For high-impact cases, especially asset theft and gas abuse, we used
Base/Solana testnets whenever supported. If only mainnet was supported, we
used at most one small-value validation transaction before stopping and
reporting the issue.
\emph{(4) Bounded high-impact validation.}
We did not conduct gas-drain experiments, availability-degrading load
tests, actual fund theft, or unpaid access to third-party services. Service
denial and gas abuse were checked only to determine whether the risk
pattern existed, and were reported as high-risk evidence when not fully
exercised. Free-shopping tests used our own server-side integration and
accounts. For asset theft, we confirmed that the system could be induced to
submit a token approval transaction, but did not perform any subsequent
transfer.


\noindent\textbf{Decision to Publish.}
We proceeded with the research because the expected security benefits to the ecosystem are substantial and because we can bound direct impacts on facilitators, services and users.
We believe publication provides net positive impact because the identified risks
stem from systemic design and integration patterns that are likely to recur
across deployments. We therefore prioritize actionable guidance for facilitators
and server implementations, including security rules that make verification and
settlement outcomes consistent and reduce the attack surface. To limit misuse, we redact or delay sensitive exploit details as needed and align artifact release with responsible disclosure and patch availability.

\section*{Open Science}
We support open and reproducible research while limiting release of materials
that could enable exploitation of live payment infrastructure. 
We release artifacts through three channels: a public Zenodo record
\url{https://zenodo.org/records/20328961}, a GitHub repository for non-sensitive
artifacts and future maintenance \url{https://github.com/HexHive/x402scope},
and a restricted-access Zenodo record
\url{https://zenodo.org/records/20329070}.

\begin{enumerate}
\item\textbf{Public Zenodo record and GitHub repository.} We provide the sanitized \sysname{} framework together with measurement code, configuration files, MariaDB
  schema and ingestion scripts, SQL queries/views, and plot/table regeneration
  code. The GitHub repository
  will mirror the non-sensitive public Zenodo artifacts and support ongoing
  maintenance.

\item\textbf{Restricted-access Zenodo repository.}
The restricted record supports artifact evaluation and follow-on research
without broadly publishing exploit-enabling details. 
It contains the full \sysname{} codebase, including mutation and deployed contract code as well as attack PoCs. It also includes the per-facilitator SR pass/fail matrix, HTTP and on-chain evidence logs for the \violateruleNum{} SR violations, adjudication records linking those violations to the \bugCount{} exploitable instances, and the target and configuration files used in the evaluation.
\end{enumerate}

To prevent misuse, we do not release weaponized exploit code or unfixed
vulnerability details in the public record. Exploit-enabling checks, mutation
payload generators, evaluation-only ERC-1271/6492 contracts, and live target
identifiers are removed from public artifacts and placed only in the
restricted bundle when needed to substantiate core findings. Access requests
will be reviewed for legitimate research or evaluation and commitment
to responsible use. 


\bibliographystyle{plainurl}
\bibliography{ref}

@inproceedings{pavlides2025more,
  title={More is Less: Extra Features in Contactless Payments Break Security},
  author={Pavlides, George and Clee, Anna and Boureanu, Ioana and Chothia, Tom},
  booktitle={Proceedings of the 34th USENIX Security Symposium (USENIX Security '25)},
  year={2025},
  pages={7977--7996},
  url={https://www.usenix.org/conference/usenixsecurity25/presentation/pavlides}
}

@inproceedings{olivier2025pays,
  title={Who Pays Whom? {Anonymous} {EMV-Compliant} Contactless Payments},
  author={Olivier-Anclin, Charles and Boureanu, Ioana and Chen, Liqun and Newton, Christopher and Chothia, Tom and Clee, Anna and Kokkinis, Andreas and Lafourcade, Pascal},
  booktitle={Proceedings of the 34th USENIX Security Symposium (USENIX Security '25)},
  year={2025},
  url={https://www.usenix.org/conference/usenixsecurity25/presentation/olivier-anclin}
}

@inproceedings{coppola2024pure,
  title={{PURE}: Payments with {UWB} {RElay-protection}},
  author={Coppola, Daniele and Camurati, Giovanni and Anliker, Claudio and Hofmeier, Xenia and Schaller, Patrick and Basin, David and Capkun, Srdjan},
  booktitle={Proceedings of the 33rd USENIX Security Symposium (USENIX Security '24)},
  pages={4553--4569},
  year={2024},
  url = {https://www.usenix.org/conference/usenixsecurity24/presentation/coppola},
}

@inproceedings{bursuc2023provably,
  title={Provably unlinkable smart card-based payments},
  author={Bursuc, Sergiu and Horne, Ross and Mauw, Sjouke and Yurkov, Semen},
  booktitle={Proceedings of the 2023 ACM SIGSAC Conference on Computer and Communications Security (CCS '23)},
  pages={1392--1406},
  year={2023},
  doi={10.1145/3576915.3623109}
}

@inproceedings{boureanu2020security,
  title={Security analysis and implementation of relay-resistant contactless payments},
  author={Boureanu, Ioana and Chothia, Tom and Debant, Alexandre and Delaune, St{\'e}phanie},
  booktitle={Proceedings of the 2020 ACM SIGSAC Conference on Computer and Communications Security (CCS '20)},
  pages={879--898},
  year={2020},
  doi={10.1145/3372297.3417235}
}

@inproceedings{rahaman2019security,
  title={Security certification in payment card industry: Testbeds, measurements, and recommendations},
  author={Rahaman, Sazzadur and Wang, Gang and Yao, Danfeng},
  booktitle={Proceedings of the 2019 ACM SIGSAC Conference on Computer and Communications Security (CCS '19)},
  pages={481--498},
  year={2019},
  doi={10.1145/3319535.3363195}
}

@inproceedings{do2022formal,
  title={A Formal Security Analysis of the {W3C} {Web} Payment {APIs}: Attacks and Verification},
  author={Do, Quoc Huy and Hosseyni, Pedram and K{\"u}sters, Ralf and Schmitz, Guido and Wenzler, Nils and W{\"u}rtele, Tim},
  booktitle={2022 IEEE Symposium on Security and Privacy (SP '22)},
  pages={215--234},
  year={2022},
  organization={IEEE},
  doi={10.1109/SP46214.2022.9833681}
}

@inproceedings{lou2021messy,
  title={Messy states of wiring: Vulnerabilities in emerging personal payment systems},
  author={Lou, Jiadong and Yuan, Xu and Zhang, Ning},
  booktitle={Proceedings of the 30th USENIX Security Symposium (USENIX Security '21)},
  pages={3273--3289},
  year={2021},
  url = {https://www.usenix.org/conference/usenixsecurity21/presentation/lou},
}

@inproceedings{kumar2020security,
  title={Security analysis of unified payments interface and payment apps in {India}},
  author={Kumar, Renuka and Kishore, Sreesh and Lu, Hao and Prakash, Atul},
  booktitle={Proceedings of the 29th USENIX Security Symposium (USENIX Security '20)},
  pages={1499--1516},
  year={2020},
  url = {https://www.usenix.org/conference/usenixsecurity20/presentation/kumar}
}

@inproceedings{chen2019devils,
  title={Devils in the guidance: predicting logic vulnerabilities in payment syndication services through automated documentation analysis},
  author={Chen, Yi and Xing, Luyi and Qin, Yue and Liao, Xiaojing and Wang, XiaoFeng and Chen, Kai and Zou, Wei},
  booktitle={Proceedings of the 28th USENIX Security Symposium (USENIX Security '19)},
  pages={747--764},
  year={2019},
  url = {https://www.usenix.org/conference/usenixsecurity19/presentation/chen-yi},
}

@inproceedings{abi2021doing,
  title={Doing good by fighting fraud: Ethical anti-fraud systems for mobile payments},
  author={Abi Din, Zainul and Venugopalan, Hari and Lin, Henry and Wushensky, Adam and Liu, Steven and King, Samuel T},
  booktitle={2021 IEEE Symposium on Security and Privacy (SP '21)},
  pages={1623--1640},
  year={2021},
  organization={IEEE},
  doi={10.1109/SP40001.2021.00100}
}

@article{birch2025agentic,
  title={Agentic commerce and payments: Exploring the implications of robots paying robots},
  author={Birch, David GW and Gamble, Debbie},
  journal={Journal of Payments Strategy \& Systems},
  volume={19},
  number={1},
  pages={72--84},
  year={2025},
  publisher={Henry Stewart Publications},
  doi={10.69554/NGEA2302}
}

@article{rothschild2025agentic,
  title={The Agentic Economy},
  author={Rothschild, David M and Mobius, Markus and Hofman, Jake M and Dillon, Eleanor W and Goldstein, Daniel G and Immorlica, Nicole and Jaffe, Sonia and Lucier, Brendan and Slivkins, Aleksandrs and Vogel, Matthew},
  journal={arXiv preprint arXiv:2505.15799},
  year={2025},
  url={https://arxiv.org/abs/2505.15799}
}

@article{sapkota2025ai,
  title={{AI} agents vs. agentic {AI}: A conceptual taxonomy, applications and challenges},
  author={Sapkota, Ranjan and Roumeliotis, Konstantinos I and Karkee, Manoj},
  journal={arXiv preprint arXiv:2505.10468},
  year={2025},
  url={https://arxiv.org/abs/2505.10468}
}

@techreport{teamhuman,
  title       = {From Human-Centric to Agent-Native: Building Trustless Payment Infrastructure for Agentic {AI}},
  author      = {Shi, Scott and Cheng, Zerui and Xi, Chen and Huang, Yi and Li, Lyon and Marwaha, Uddhav and Weber, David and Zhang, Chi},
  institution = {Kite AI},
  year        = {2025},
  month       = {Oct},
  day         = {30},
  type        = {Whitepaper},
  url         = {https://www.zerui-cheng.com/uploads/Kite_whitepaper.pdf},
  note        = {Accessed: 2026-01-17}
}

@article{shi2025sybil,
  title={Sybil-Resistant Service Discovery for Agent Economies},
  author={Shi, David and Joo, Kevin},
  journal={arXiv preprint arXiv:2510.27554},
  year={2025},
  url={https://arxiv.org/abs/2510.27554}
}

@misc{coinbase_x402,
  title        = {The {Internet-native} Payment Protocol},
  author       = {{Coinbase}},
  howpublished = {\url{https://www.coinbase.com/developer-platform/products/x402}},
  note         = {Accessed: 2026-01-17}
}

@misc{x402scan,
  title        = {The x402 Analytics Dashboard and Block Explorer},
  author       = {{x402scan.com}},
  howpublished = {\url{https://www.x402scan.com/}},
  note         = {Accessed: 2026-01-17}
}

@misc{google_a2a_x402,
  title        = {{The Agent-to-Agent (A2A) protocol x402 Extension}},
  author       = {{Google}},
  howpublished = {\url{https://github.com/google-agentic-commerce/a2a-x402}},
  note         = {Accessed: 2026-01-17}
}

@misc{cloudflare_agents,
  title        = {Cloudflare Agents},
  author       = {{Cloudflare}},
  howpublished = {\url{https://github.com/cloudflare/agents}},
  note         = {Accessed: 2026-01-17}
}

@misc{yakovenko2018solana,
  author       = {Anatoly Yakovenko},
  title        = {{Solana}: A New Architecture for a High Performance Blockchain v0.8.13},
  howpublished = {Whitepaper},
  year         = {2018},
  url          = {https://solana.com/solana-whitepaper.pdf},
  note         = {Accessed: 2026-01-17}
}

@misc{coingecko,
  title        = {Cryptocurrency Prices by Market Cap},
  howpublished = {\url{https://www.coingecko.com/}},
  note         = {Accessed: 2026-01-01},
  organization = {CoinGecko}
}

@misc{SolDoc,
  title        = {{Solana Documentation: Accounts}},
  howpublished = {\url{https://docs.solana.com/developing/programming-model/accounts#rent-exemption}},
  note         = {Accessed: 2026-01-25},
  author = {Solana Foundation}
}

@misc{x402secure,
  title        = {x402-secure: Secure Every Agent Payment on x402},
  howpublished = {\url{https://www.x402secure.com/}},
  note         = {Accessed: 2026-01-25},
  organization = {t54 labs}
}

@misc{hashed_official_x402_analytics,
  title        = {x402 Analytics},
  author       = {{hashed\_official}},
  howpublished = {Dune dashboard},
  url          = {https://dune.com/hashed_official/x402-analytics},
  note         = {Accessed 2026-02-02}
}

@inproceedings{augusto2024sok,
  title={{SoK}: Security and privacy of blockchain interoperability},
  author={Augusto, Andr{\'e} and Belchior, Rafael and Correia, Miguel and Vasconcelos, Andr{\'e} and Zhang, Luyao and Hardjono, Thomas},
  booktitle={2024 IEEE Symposium on Security and Privacy (SP '24)},
  pages={3840--3865},
  year={2024},
  organization={IEEE},
  doi={10.1109/SP54263.2024.00255}
}

@inproceedings{thyagarajan2022universal,
  title={Universal atomic swaps: Secure exchange of coins across all blockchains},
  author={Thyagarajan, Sri AravindaKrishnan and Malavolta, Giulio and Moreno-Sanchez, Pedro},
  booktitle={2022 IEEE Symposium on Security and Privacy (SP '22)},
  pages={1299--1316},
  year={2022},
  organization={IEEE},
  doi={10.1109/SP46214.2022.9833731}
}

@inproceedings{luo2024towards,
  title={Towards Automatic Discovery of Denial of Service Weaknesses in Blockchain Resource Models},
  author={Luo, Feng and Lin, Huangkun and Li, Zihao and Luo, Xiapu and Luo, Ruijie and He, Zheyuan and Song, Shuwei and Chen, Ting and Luo, Wenxuan},
  booktitle={Proceedings of the 2024 ACM SIGSAC Conference on Computer and Communications Security (CCS '24)},
  pages={1016--1030},
  year={2024},
  doi={10.1145/3658644.3690329}
}

@inproceedings{wang2024smartinv,
  title={{SmartInv}: Multimodal learning for smart contract invariant inference},
  author={Wang, Sally Junsong and Pei, Kexin and Yang, Junfeng},
  booktitle={2024 IEEE Symposium on Security and Privacy (SP '24)},
  pages={2217--2235},
  year={2024},
  organization={IEEE},
  doi={10.1109/SP54263.2024.00126}
}

@inproceedings{sendner2024large,
  title={Large-scale study of vulnerability scanners for {Ethereum} smart contracts},
  author={Sendner, Christoph and Petzi, Lukas and Stang, Jasper and Dmitrienko, Alexandra},
  booktitle={2024 IEEE Symposium on Security and Privacy (SP '24)},
  pages={2273--2290},
  year={2024},
  organization={IEEE},
  doi={10.1109/SP54263.2024.00230}
}

@inproceedings{li2024enhancing,
  title={Demo: Enhancing Smart Contract Security Comprehensively through Dynamic Symbolic Execution},
  author={Li, Zhaoxuan and Zhao, Ziming and Li, Wenhao and Zhang, Rui and Xue, Rui and Lu, Siqi and Zhang, Fan},
  booktitle={Proceedings of the 2024 ACM SIGSAC Conference on Computer and Communications Security (CCS '24)},
  pages={5072--5074},
  year={2024},
  doi={10.1145/3658644.3691365}
}

@inproceedings{smolka2023fuzz,
  title={Fuzz on the beach: Fuzzing {Solana} smart contracts},
  author={Smolka, Sven and Giesen, Jens-Rene and Winkler, Pascal and Draissi, Oussama and Davi, Lucas and Karame, Ghassan and Pohl, Klaus},
  booktitle={Proceedings of the 2023 ACM SIGSAC Conference on Computer and Communications Security (CCS '23)},
  pages={1197--1211},
  year={2023},
  doi={10.1145/3576915.3623178}
}

@inproceedings{babel2023lanturn,
  title={Lanturn: Measuring economic security of smart contracts through adaptive learning},
  author={Babel, Kushal and Javaheripi, Mojan and Ji, Yan and Kelkar, Mahimna and Koushanfar, Farinaz and Juels, Ari},
  booktitle={Proceedings of the 2023 ACM SIGSAC Conference on Computer and Communications Security (CCS '23)},
  pages={1212--1226},
  year={2023},
  doi={10.1145/3576915.3623204}
}

@inproceedings{pan2025collisionrepair,
  title={{CollisionRepair}: {First-Aid} and Automated Patching for Storage Collision Vulnerabilities in Smart Contracts},
  author={Pan, Yu and Han, Wanjing and Duan, Yue and Zhang, Mu},
  booktitle={Proceedings of the 34th USENIX Security Symposium (USENIX Security '25)},
  pages={4035--4052},
  year={2025},
  url={https://www.usenix.org/conference/usenixsecurity25/presentation/pan-yu}
}

@inproceedings{sun2024all,
  title={All your tokens are belong to us: Demystifying address verification vulnerabilities in solidity smart contracts},
  author={Sun, Tianle and He, Ningyu and Xiao, Jiang and Yue, Yinliang and Luo, Xiapu and Wang, Haoyu},
  booktitle={Proceedings of the 33rd USENIX Security Symposium (USENIX Security '24)},
  pages={3567--3584},
  year={2024},
  url = {https://www.usenix.org/conference/usenixsecurity24/presentation/sun-tianle},
}

@inproceedings{wang2023automated,
  title={Automated inference on financial security of {Ethereum} smart contracts},
  author={Wang, Wansen and Huang, Wenchao and Meng, Zhaoyi and Xiong, Yan and Miao, Fuyou and Fang, Xianjin and Tu, Caichang and Ji, Renjie},
  booktitle={Proceedings of the 32nd USENIX Security Symposium (USENIX Security '23)},
  pages={3367--3383},
  year={2023},
  url={https://www.usenix.org/conference/usenixsecurity23/presentation/wang-wansen}
}

@inproceedings{sendner2023smarter,
  title={Smarter Contracts: Detecting Vulnerabilities in Smart Contracts with Deep Transfer Learning.},
  author={Sendner, Christoph and Chen, Huili and Fereidooni, Hossein and Petzi, Lukas and K{\"o}nig, Jan and Stang, Jasper and Dmitrienko, Alexandra and Sadeghi, Ahmad-Reza and Koushanfar, Farinaz},
  booktitle={Proceedings of the 30th Network and Distributed System Security Symposium (NDSS '23)},
  year={2023},
  url          = {https://www.ndss-symposium.org/ndss-paper/smarter-contracts-detecting-vulnerabilities-in-smart-contracts-with-deep-transfer-learning/},
}

@ARTICLE{zhang2024demystifying,
  author={Zhang, Jiashuo and Chen, Jiachi and Shen, Yiming and Zhang, Tao and Wang, Yanlin and Chen, Ting and Gao, Jianbo and Chen, Zhong},
  journal={IEEE Transactions on Software Engineering}, 
  title={When Crypto Fails: Demystifying Cryptographic Defects in {Ethereum} Smart Contracts}, 
  year={2025},
  volume={51},
  number={5},
  pages={1381-1398},
  doi={10.1109/TSE.2025.3551776}}

@misc{coinbase_x402_spec,
  author       = {{Coinbase}},
  title        = {{x402 Specification}},
  howpublished = {\url{https://github.com/coinbase/x402/blob/b4464ce/specs/x402-specification.md}},
  year         = {2026},
  note         = {Accessed: 2026-01-17}
}

@misc{erc6492,
  author       = {Georgiev, Ivo and Aguilar, Agustin},
  title        = {{ERC-6492}: Signature Validation for Predeploy Contracts},
  howpublished = {Ethereum Improvement Proposals},
  number       = {6492},
  month        = feb,
  year         = {2023},
  url          = {https://eips.ethereum.org/EIPS/eip-6492}
}

@misc{erc3009,
  author       = {Peter Jihoon Kim and Kevin Britz and David Knott},
  title        = {{ERC-3009}: Transfer With Authorization},
  howpublished = {Ethereum Improvement Proposals},
  number       = {3009},
  note         = {Draft},
  year         = {2020},
  month = oct,
  url          = {https://eips.ethereum.org/EIPS/eip-3009}
}

@misc{solana_transfer_tokens,
  author       = {{Solana Foundation}},
  title        = {Transfer Tokens},
  howpublished = {Solana Documentation},
  url          = {https://solana.com/docs/tokens/basics/transfer-tokens},
  note         = {Accessed 2026-02-05}
}

@misc{eip712,
  author       = {Remco Bloemen and Leonid Logvinov and Jacob Evans},
  title        = {{EIP-712}: Ethereum Typed Structured Data Hashing and Signing},
  howpublished = {Ethereum Improvement Proposals},
  number       = {712},
  year         = {2017},
  url          = {https://eips.ethereum.org/EIPS/eip-712},
  note         = {Accessed 2026-02-05}
}

@misc{erc1271,
  author       = {Jacques Dafflon and Alex Beregszaszi},
  title        = {{EIP-1271}: Standard Signature Validation Method for Contracts},
  howpublished = {Ethereum Improvement Proposals},
  number       = {1271},
  year         = {2018},
  url          = {https://eips.ethereum.org/EIPS/eip-1271},
  note         = {Accessed 2026-02-05}
}

@misc{aws_builder_x402,
  title        = {Monetize Any {HTTP} Application with {x402} and {CloudFront} {Lambda@Edge}},
  author       = {{Amazon Web Services}},
  howpublished = {\url{https://builder.aws.com/content/38fLQk6zKRfLnaUNzcLPsUexUlZ/monetize-any-http-application-with-x402-and-cloudfront-lambdaedge}},
  year         = {2026},
  note         = {Accessed: 2026-02-06},
}

@misc{circle_blog_x402,
  title        = {Autonomous Payments using {Circle} Wallets, {USDC}, and x402},
  author       = {{Circle Internet Financial}},
  year         = {2025},
  howpublished = {\url{https://www.circle.com/blog/autonomous-payments-using-circle-wallets-usdc-and-x402}},
  note         = {Published September 12, 2025; accessed 2026-02-06},
}

@inproceedings{chen2025unveiling,
  title={Unveiling Security Vulnerabilities in Git Large File Storage Protocol},
  author={Chen, Yuan and Wang, Qinying and Yang, Yong and Chen, Yuanchao and Li, Yuwei and Ji, Shouling},
  booktitle={2025 IEEE Symposium on Security and Privacy (SP)},
  pages={468--485},
  year={2025},
  organization={IEEE}
}

\end{document}